\documentclass[twocolumn,english,pra]{revtex4-1}
\usepackage[T1]{fontenc}
\usepackage[utf8]{inputenc}
\usepackage{color}
\usepackage{amsmath}
\usepackage{amssymb}
\usepackage{graphicx}
\usepackage{wasysym}

\makeatletter

\newcommand{\lyxdot}{.}

\usepackage[colorlinks = true,
            linkcolor =red,
            urlcolor  =magenta,
            citecolor = blue,
            anchorcolor = blue]{hyperref}
\begingroup
\footnotetext{These authors contributed equally to this work.}

\makeatother

\usepackage{babel}
\begin{document}
\title{Enhancement of nonclassical properties of two-mode squeezed vacuum
state with postselected von Neumann measurement}
\author{Janarbek Yuanbek$^{1,2}$\textsuperscript{\textsection}, Yi-Fang
Ren$^{1}$\textsuperscript{\textsection}, Ahmad Abliz$^{2}$}
\email{aahmad@126.com}

\author{Yusuf Turek$^{1}$}
\email{yusuftu1984@hotmail.com}

\affiliation{$^{1}$School of Physics, Liaoning University, Shenyang, Liaoning
110036, China}
\affiliation{$^{2}$School of Physics and Electronic Engineering, Xinjiang Normal
University, Urumqi, Xinjiang 830054, China}
\date{\today}
\begin{abstract}
We investigate the effects of weak value amplification on the nonclassical
properties of the two-mode squeezed vacuum state. We calculate the
quadrature squeezing and second-order cross-correlation functions
to demonstrate the advantages of post-selective WMs. We also examine
the entanglement between the two modes by evaluating various entanglement
criteria, including the Cauchy-Schwartz inequality, Hillery-Zubairy
correlation, and Einstein-Podolsky-Rosen correlations. Our results
show that postselected von Neumann measurements significantly optimize
the two-mode squeezed vacuum state. Notably, the amplification effect
of weak values plays a critical role in this process. We demonstrated
that in the WM regime, as the weak values increase, the nonclassical
properties of the two-mode squeezed vacuum state such as entanglement
and correlations between the two modes become more pronounced compared
to the original state. We also propose possible implementation methods
for our approach using trapped ion systems and optical platforms.
Our enhanced two-mode squeezed vacuum state may have potential applications
in quantum information processing scenarios, such as single-photon
generation, advanced imaging techniques, quantum computation, and
long-distance quantum communication.
\end{abstract}
\maketitle

\section{Introduction\label{sec:1}}

The squeezed light field is a crucial nonclassical radiation field
in quantum optics \citep{1}. It is characterized by two quadrature
operators along distinct directions. When the fluctuations in one
quadrature operator are reduced below the level of those in a coherent
state, the fluctuations in the other quadrature become correspondingly
larger. This unique property of squeezing has enabled numerous applications
in quantum information processing, including quantum cryptography
\citep{RevModPhys.74.145}, quantum communication (e.g., teleportation)
\citep{PhysRevLett.124.060501}, quantum coding \citep{PhysRevA.51.2738},
and quantum state optimization \citep{PhysRevA.98.022115}. As the
use of squeezed states becomes increasingly prevalent, the study of
squeezing operators and squeezed states has emerged as a significant
topic in both quantum optics and quantum information \citep{Andersen_2016}.
Consequently, squeezed states have become indispensable tools in these
fields. There are various types of squeezed states, including the
squeezed vacuum state \citep{RN1932}, squeezed coherent state \citep{PhysRevLett.75.4011,PhysRevA.13.2226},
photon-added or photon-subtracted squeezed states \citep{PhysRevA.75.032104,PhysRevA.43.492}
and the two-mode squeezed vacuum (TMSV) state \citep{Riabinin_2021,Daoming2015QuantumPO}.
Among these, the TMSV state stands out for inherently possessing two
important nonclassical features: squeezing and entanglement. It can
be generated via spontaneous parametric down-conversion in a nonlinear
crystal \citep{PhysRevA.31.3093,PhysRevLett.65.1838,PhysRevLett.132.246902,PhysRevA.77.033808,PhysRevA.96.023818}.
As a prototypical two-mode Gaussian state, the TMSV state plays a
vital role in various continuous-variable (CV) quantum information
processing (QIP) protocols, such as continuous-variable quantum teleportation
\citep{PhysRevLett.80.869,PhysRevA.68.052308}, weak-field detection
\citep{PhysRevApplied.19.044062,PhysRevLett.119.043603}, and quantum
key distribution \citep{2018lA}. In recent years, there has been
a growing interest in non-Gaussian states, which are produced through
non-Gaussian operations such as photon addition and subtraction applied
to initial Gaussian states \citep{2020quantum,PRXQuantum.2.030204}.
Multimode non-Gaussian states, such as the enhanced TMSV state, exhibit
strong quantum correlations between the two modes in terms of photon
number, frequency, time, and position. Investigating these correlations
in photon-added or photon-subtracted TMSV states reveals that these
enhanced TMSV states are highly useful in a range of quantum technologies
\citep{2020quantum,PhysRevA.109.040101}, including single and multi-parameter
loss sensing \citep{PhysRevLett.107.193602,PhysRevLett.121.230801,2023},
quantum communication \citep{MasashiBan_1999,2021S,2024}, quantum
illumination \citep{PhysRevA.103.062413,PhysRevApplied.20.044001,PhysRevA.109.062440}
and quantum metrology \citep{PhysRevLett.104.103602,Ouyang:16}. However,
dealing with non-Gaussian states and operations presents additional
computational complexity compared to Gaussian states, often requiring
more intricate calculations. Therefore, there is a need for alternative,
more accessible methods to optimize these states. Specifically, to
improve the efficiency of TMSV-based applications without relying
on photon subtraction, it is essential to explore new state-optimization
schemes that further enhance the nonclassical properties of the TMSV
state.

Quantum state optimization is a crucial method for enhancing the efficiency
of quantum information processing, and its effectiveness depends on
quantum measurement and control techniques \citep{2014Quantum}. The
theory of quantum WM (WM), introduced by Aharonov, Albert, and Vaidman
in the late 1980s \citep{PhysRevLett.60.1351}, not only extends the
scope of traditional measurement theory but also opens new possibilities
for quantum state engineering \citep{PhysRevA.105.022608}. One key
feature distinguishing WM from strong measurement is that the weak
value of an observable can exceed the typical range of its eigenvalues
\citep{Aharonov2005}. This characteristic, an amplification effect
for weak signals, has proven highly effective in solving various challenges
in physics and related fields. We refer readers to recent reviews
in the field for a detailed overview \citep{70,RN2035,RN2066,71}.
In WM, only the first-order evolution of the unitary operator is important,
as the interaction strength between the system and the measuring device
is minimal. However, to bridge the gap between weak and strong measurements,
analyze postselected WM feedback, and interpret experimental results
obtained under non-ideal conditions, it is essential to consider the
full-order evolution of the unitary operator. This type of measurement
is known as postselected von Neumann measurement. The signal amplification
properties inherent in WM can optimize quantum states. Recently, there
has been considerable interest in state optimization via postselected
von Neumann measurements, particularly where the pointer states are
Gaussian states \citep{oh2019optimal,PhysRevA.67.052311}, Hermite-Gaussian
or Laguerre-Gaussian states, and nonclassical states \citep{PhysRevA.92.022109,PhysRevA.105.022210}.
In particular, recent studies have examined the advantages of using
nonclassical pointer states to improve the precision of postselected
measurements \citep{PhysRevA.92.022109,Turek_2020,2021Single}. These
studies have demonstrated that postselected von Neumann measurement
can significantly alter photon statistics and radiation fields' phase
space distributions, especially in regimes of anomalously large weak
values. However, to our knowledge, researchers have only applied postselected
von Neumann measurement-based quantum state optimization to single-mode
radiation fields, and its potential in multimode cases remains unexplored.
As mentioned earlier, the TMSV state is a representative two-mode
radiation field. Thus, one might ask whether it is possible to apply
postselected von Neumann measurement-based state optimization to enhance
the nonclassical characteristics of the TMSV state. Our answer to
this question is affirmative.

In this work, we apply postselected measurement techniques to enhance
the nonclassical properties of the TMSV state. Specifically, we use
the postselected WM procedure to one mode of the TMSV state, treating
the TMSV state as the pointer and its polarization degree of freedom
as the measured system. After obtaining a normalized final pointer
state, modified by the weak value of the measured observable, we analyze
the resulting nonclassical characteristics of the output TMSV state.
We present analytical expressions and numerical results, examining
properties such as quadrature squeezing, second-order cross-correlation
functions, and various entanglement criteria for two-mode fields.
Our findings demonstrate a significant enhancement of the intrinsic
quantum properties of the TMSV state after postselected WM, thanks
to weak value amplification. The method proposed in this paper offers
an alternative approach to improve the efficiency of TMSV-state-based
quantum information processing and other technical applications.

We organized the paper as follows: In Sec. \ref{sec:2}, we introduce
the main concepts of our scheme and derive the final pointer state
after postselected measurement, which serves as the foundation for
our analysis. To clearly illustrate the effects of postselected von
Neumann measurement on the relevant characteristics of the TMSV state,
we first examine the quadrature squeezing of the final pointer state
in Sec. \ref{sec:3}. In Sec. \ref{sec:4}, we analyze the quantum
statistics of the output pointer state, focusing on the Mandel-$Q$
parameter and second-order correlation functions. To further evaluate
the impact of postselected von Neumann measurement on one mode of
the TMSV state, we investigate various entanglement criteria, including
the Cauchy-Schwartz inequality, Hillery-Zubairy correlations, and
Einstein-Podolsky-Rosen (EPR) correlations in Sec. \ref{sec:5}. Sec.
\ref{sec:6} provides a brief analysis of the degree of deviation
from the initial pointer state, quantified by the fidelity between
the input and output pointer states to assess the changes to the pointer
state after the postselected von Neumann measurement. Sec. \ref{sec:7}
presents potential experimental implementations of our theoretical
model using trapped ion systems and optical platforms. Sec. \ref{sec:8}
discusses the broader implications of our findings, and we conclude
the paper in Sec. \ref{sec:9} with final remarks and suggestions
for future work. Throughout this paper, we set $\hbar=1$.

\section{\label{sec:2}MODEL SETUP}

Let us consider a TMSV state, represented as $\vert\phi\rangle=\vert\xi\rangle=\hat{S}(\xi)\vert0\rangle_{a}\vert0\rangle_{b}=\hat{S}(\xi)\vert0,0\rangle$,
where $\hat{S}(\xi)=e^{\xi a^{\dagger}b^{\dagger}-\xi^{\ast}ab}$
is the two-mode squeezing operator. Here, $a$ ($a^{\dagger}$) and
$b$ ($b^{\dagger}$) are the annihilation (creation) operators for
the two bosonic modes, satisfying the commutation relations $[\hat{a},\hat{a}^{\dagger}]=[\hat{b},\hat{b}^{\dagger}]=1$
and $[\hat{a},\hat{b}]=0$. Here, $\xi=\lambda e^{i\theta}$ and $\lambda$
represents the squeezing parameter, with $0\leq\lambda<\infty$ and
$0\leq\theta\leq2\pi$. Various platforms allow generating the TMSV
state, including optical setups \citep{2005,2018,REN2019106,PhysRevResearch.3.033095,Riabinin_2021,PhysRevLett.132.246902},
Josephson traveling wave parametric amplifiers \citep{PhysRevLett.107.113601,2017,fasolo2021,10.1063/5.0064892,PhysRevLett.128.153603}
, atomic \citep{PhysRevResearch.3.033095,PhysRevLett.131.193601}
and ion trap systems \citep{ZENG2002427,PhysRevA.104.032609}. 

As discussed in the introduction, the TMSV state exhibits intrinsic
quantum correlations between its two modes, making it highly applicable
to numerous studies. This work focuses on optimizing the TMSV state
using weak value amplification, a technique designed to amplify weak
signals. In the standard framework of postselected WM, we treat the
TMSV state as the pointer and its polarization as the measured system.
For simplicity, we assume that the interaction between pointer and
measured system occurs only on one mode, specifically, the $a$ mode
of the TMSV beam and its polarization. A von Neumann-type Hamiltonian
describes this kind of interaction can be written as 

\begin{equation}
\hat{H}_{int}=g\hat{\sigma}_{x}\otimes\hat{P}_{x}.\label{eq:1}
\end{equation}
Here, $g$ represents the interaction coupling parameter between the
pointer and the measured system, $\hat{\sigma}_{x}=\vert D\rangle\langle D\vert-\vert A\rangle\langle A\vert$,
where $\vert D\rangle=\frac{1}{\sqrt{2}}\left(\vert H\rangle+\vert V\rangle\right)$
and $\vert A\rangle=\frac{1}{\sqrt{2}}\left(\vert H\rangle-\vert V\rangle\right)$
represent diagonal and anti-diagonal polarization of the beam, respectively.
In the interaction Hamiltonian above, $\hat{P}_{x}$ represents the
momentum observable of the pointer. In terms of annihilation and creation
operators, it is given by: 
\begin{equation}
\hat{P}_{x}=\frac{i}{2\sigma}\left(\hat{a}^{\dagger}-\hat{a}\right),\label{eq:2}
\end{equation}
where $\sigma=\sqrt{1/2m\omega}$ is the width of the Gaussian ground
state, which depends on the mass of the pointer, $m$, and the frequency,
$\omega$, at which the system oscillates. We now assume an initial
state of the composite system prepared in the form 
\begin{equation}
\vert\Psi_{in}\rangle=\vert\psi_{i}\rangle\otimes\vert\phi\rangle,\label{eq:3}
\end{equation}
where 
\begin{equation}
\vert\psi_{i}\rangle=\cos\frac{\alpha}{2}\vert H\rangle+e^{i\delta}\sin\frac{\alpha}{2}\vert V\rangle\label{eq:4}
\end{equation}
and $\vert\phi\rangle$ are the initial states of the measured system
and pointer, respectively. Here, $\delta\in[0,2\pi]$ and $\alpha\in[0,\pi)$.
In the optical lab, we prepare this initial state by passing the beam
through a quarter-wave and half-wave plates, with their optical axes
set at appropriate angles. The joint state expressed in Eq. (\ref{eq:3})
evolves by means of the time evolution operator, $U(t)=\exp\left(-i\int_{0}^{t}H_{int}d\tau\right)$,
as 
\begin{eqnarray}
\vert\Psi_{evol}\rangle & = & \exp\left(-i\int_{0}^{t}\hat{H}_{int}d\tau\right)\vert\Psi_{in}\rangle\nonumber \\
 & = & \frac{1}{2}\left[(\mathbb{I}+\hat{\sigma}_{x})\otimes\hat{D}\left(\frac{s}{2}\right)+(\mathbb{I}-\hat{\sigma}_{x})\otimes\hat{D}\left(-\frac{s}{2}\right)\right]\nonumber \\
 &  & \times\vert\psi_{i}\rangle\otimes\vert\phi\rangle.
\end{eqnarray}
where $s=gt/\sigma$, $\mathbb{I}$ is $2\times2$ unit matrix operator
and $\hat{D}(s/2)=e^{s/2(\hat{a}^{\dagger}-\hat{a})}$ is the displacement
operator. It is important to note that $s$ characterizes the measurement
strength in this study. Therefore, when the coupling between the system
and the pointer is weak (strong), the measurement is referred to as
weak (strong) for $s\ll1$ ($s\gg1$). In this context, we can refer
to $s$ as the \textit{transition parameter} of our measurement model.
Finally, by postselecting the measured system state $\vert\psi_{f}\rangle=\vert H\rangle$
on to $\vert\Psi_{evol}\rangle$, we get the final state of the pointer,
expressed as 

\begin{equation}
\vert\Psi\rangle=\frac{\kappa}{2}\left[\left(1+\langle\hat{\sigma}_{x}\rangle_{w}\right)\hat{D}\left(\frac{s}{2}\right)+\left(1-\langle\hat{\sigma}_{x}\rangle_{w}\right)\hat{D}\left(-\frac{s}{2}\right)\right]\vert\phi\rangle.\label{eq:Psi}
\end{equation}
Here, $\kappa$ is the normalization coefficient defined as 
\begin{align}
\kappa & =\sqrt{2}\left[1+\vert\langle\hat{\sigma_{x}}\rangle_{w}\vert^{2}+(1-\vert\langle\hat{\sigma}_{x}\rangle_{w}\vert^{2})\beta\right]^{-\frac{1}{2}}\label{eq:7}
\end{align}
with $\beta=\exp\left[-\frac{s^{2}\cosh(2\lambda)}{2}\right]$, and
$\langle\hat{\sigma}_{x}\rangle_{w}$ is the weak value of the system
observable $\hat{\sigma}_{x}$ defined by 
\begin{equation}
\langle\hat{\sigma}_{x}\rangle_{w}=\frac{\langle\psi_{f}\vert\hat{\sigma}_{x}\vert\psi_{i}\rangle}{\langle\psi_{f}\vert\psi_{i}\rangle}=e^{i\delta}\tan\frac{\alpha}{2}.\label{eq:8}
\end{equation}
We can obtain this weak value in a measurement with a postselection
probability of $P_{s}=\vert\langle\psi_{f}\vert\psi_{i}\rangle\vert^{2}=\cos^{2}(\frac{\alpha}{2})$.
In the above calculation of the normalization coefficient $\kappa$,
we used the following expressions \citep{Agarwal2013}

\begin{align}
\hat{S}^{\dagger}\left(\xi\right)\hat{a}\hat{S}\left(\xi\right) & =\hat{a}\cosh\lambda+\hat{b}^{\dagger}e^{i\theta}\sinh\lambda,\\
\hat{S}^{\dagger}\left(\xi\right)\hat{b}\hat{S}\left(\xi\right) & =\hat{b}\cosh\lambda+\hat{a}^{\dagger}e^{i\theta}\sinh\lambda,
\end{align}
and 
\begin{equation}
\hat{D}(t)\hat{S}(\xi)=\hat{S}(\xi)\hat{D}(c),\label{eq:11}
\end{equation}
where $c=t\cosh\lambda+t^{\ast}e^{i\phi}\sinh\lambda$. Here, $\hat{D}(t)$
and $\hat{D}(c)$ represent the displacement operators as mentioned
above. 

\begin{figure}
\begin{centering}
\includegraphics[width=8cm]{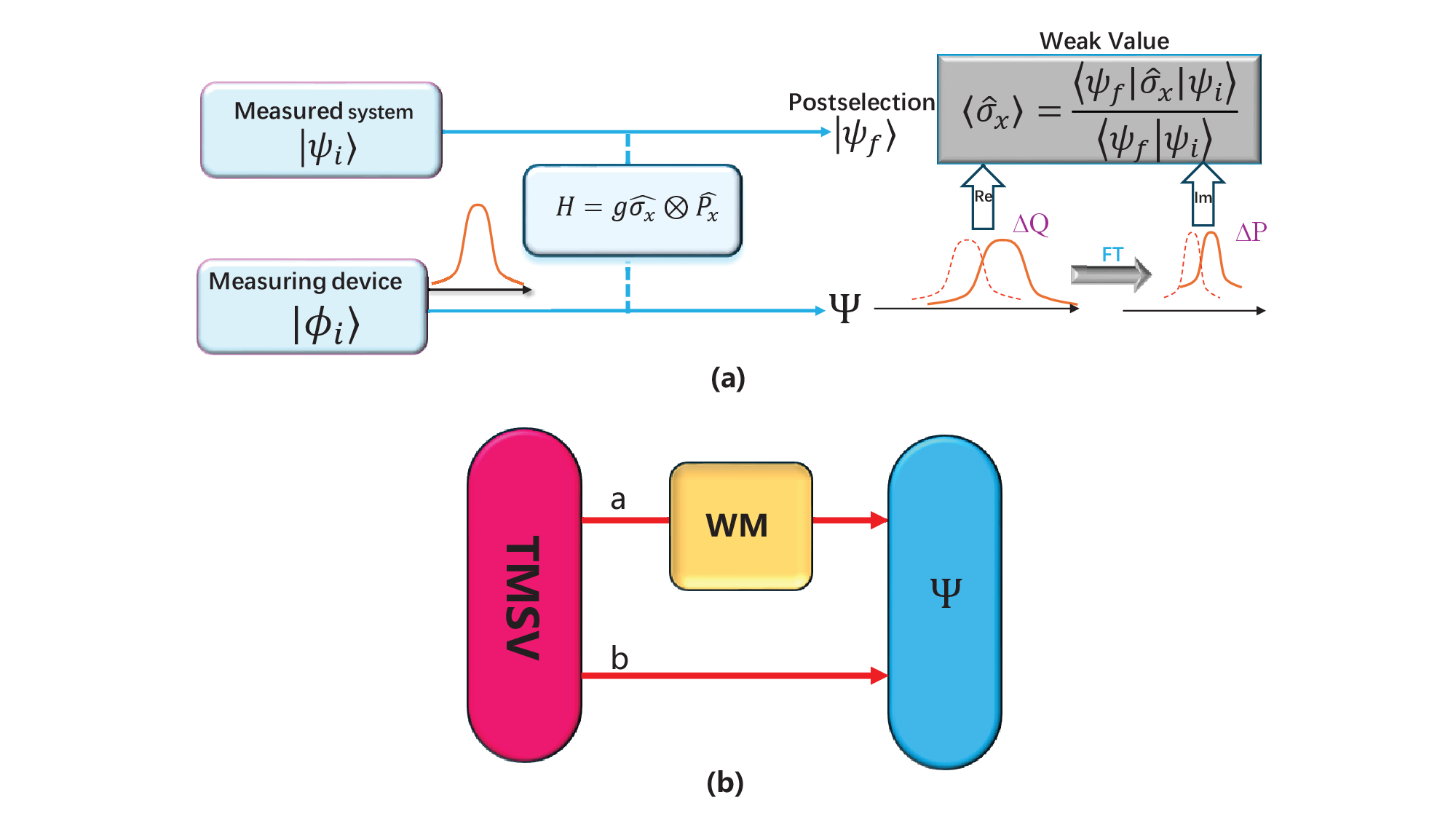}
\par\end{centering}
\caption{\label{fig:1}(a) Schematic diagram of weak measurement (WM) theory.
The standard process of WM involves four steps: (i) The initial state
of the measured system is prepared in $\vert\psi_{i}\rangle$, while
the measuring device is in the initial state $\vert\phi_{i}\rangle$.
(ii) A weak interaction occurs between the measured system and the
measuring device, during which the composite system evolves. In our
model, the Hamiltonian $H=g\,\hat{\sigma}_{x}\otimes\hat{P}_{x}$
describes the weak interaction, where a coupling constant $g$ characterizes
the bilinear coupling. (iii) After some evolution, the entire system
is projected onto the postselected state $\vert\psi_{f}\rangle$ of
the measured system. This state extracts the desired values of the
system's observable. Postselection chooses a specific subensemble
of the samples before the final measurement. (iv) The measurement
result is read out through the shifts in the pointer. In the postselected
weak measurement process, the observable value, expressed as a function
of the weak value, has its real ($\Re$) and imaginary ($\Im$) parts
read out from the shifts in position and momentum of the measuring
device, respectively. The Fourier method (FT) transforms the position
space into momentum. (b) Schematic setup for preparing the state $\vert\Psi\rangle$
via postselected von Neumann measurement.}
\end{figure}

Equation (\ref{eq:Psi}) represents the final state of the pointer
after the postselected von Neumann measurement. The weak value given
in Eq. (\ref{eq:8}) can exceed the normal range of the observable
$\hat{\sigma}_{x}$, and it can even take on complex values considering
$\delta\neq0$. As mentioned, anomalously large weak values can amplify
subtle system information and optimize quantum states. Next, we investigate
the effects of postselected von Neumann measurements and the anomalous
weak values of the measured system's observable on the inherent properties
of the TMSV state $\vert\phi\rangle$. Fig. 1 presents the schematics
of the postselected WM procedure and the setup for preparing the WM-enhanced
TMSV state.

Before beginning our investigation, there are two key points to mention:
(i) the specific weak values $\langle\hat{\sigma}_{x}\rangle_{w}$
used in the numerical analysis, where, we consider real weak values
by setting $\delta=0$, and we analyze five weak values corresponding
to $\alpha=\pi/9$, $3\pi/9$, $5\pi/9$, $7\pi/9$, and $8\pi/9$,
respectively. From Eq. (\ref{eq:8}), the weak values are as follows:
$\langle\hat{\sigma}_{x}\rangle_{w}=0.176$ for $\alpha=\pi/9$, $\langle\hat{\sigma}_{x}\rangle_{w}=0.577$
for $\alpha=\pi/3$, $\langle\hat{\sigma}_{x}\rangle_{w}=1.192$ for
$\alpha=5\pi/9$, $\langle\hat{\sigma}_{x}\rangle_{w}=2.747$ for
$\alpha=7\pi/9$, and $\langle\hat{\sigma}_{x}\rangle_{w}=5.671$
for $\alpha=8\pi/9$. In optical experiment, these angles $\alpha$
can be accomplished by adjusting the corresponding optical elements,
as introduced in Sec. VII of this paper. (ii) The value of the coupling
strength parameter $s$. As previously mentioned, $s=g\,t/\sigma$
is a dimensionless parameter determining the measurement regime. The
value of $s$ can be controlled experimentally in three ways, corresponding
to adjustments in $g$, $t$, and $\sigma$. Here, $g$ and $t$ represent
the coupling strength and interaction time between the pointer (measuring
device) and the measured system, respectively, while $\sigma$ is
the width of the wavepacket, which determines the sensitivity of the
pointer. As explored in experimental work \citep{Nature2020Yi}, the
most straightforward way to adjust the coupling strength parameter
$s$ is by tuning the interaction time $t$. In the following analysis,
we assume that changes in $s$ are due to variations in $t$, while
$g$ and $\sigma$ remain constant.

\section{The effects on quadrature squeezing \label{sec:3}}

In continuous-variable quantum optics, quadrature squeezing and non-Gaussian
entanglement are two of the most important characteristics of nonclassical
radiation fields, essential for implementing many quantum computation
and communication protocols \citep{2017A}. In this section, we examine
the effects of postselected von Neumann measurement on the quadrature
squeezing of the TMSV state. For a single-mode radiation field, squeezing
is the reduction of the quadrature variance below the shot noise level,
i.e., $\triangle^{2}X_{\vartheta}<\frac{1}{4}$. Here, $X_{\vartheta}=\frac{1}{2}\left(e^{-i\vartheta}\hat{a}+e^{i\vartheta}\hat{a}^{\dagger}\right)$
represents the quadrature operator with phase $\vartheta$, and its
variance is given by $\triangle^{2}X_{\vartheta}=\langle X_{\vartheta}^{2}\rangle-\langle X_{\vartheta}\rangle^{2}$.
Similarly, for two-mode squeezing between modes $a$ and $b$ of the
TMSV state, we define the joint quadrature operators as \citep{SCHNABEL20171}

\begin{align}
\hat{F}_{1} & =\frac{1}{\sqrt{8}}\left[e^{-i\vartheta}\left(\hat{a}+\hat{b}\right)+e^{i\vartheta}\left(\hat{a}^{\dagger}+\hat{b}^{\dagger}\right)\right],\label{eq:SQ-1}\\
\hat{F}_{2} & =\frac{1}{\sqrt{8}i}\left[e^{-i\vartheta}\left(\hat{a}+\hat{b}\right)-e^{i\vartheta}\left(\hat{a}^{\dagger}+\hat{b}^{\dagger}\right)\right].\label{eq:SQ-2}
\end{align}
These operators satisfy the commutation relation $[\hat{F}_{1},\hat{F}_{2}]=\frac{i}{2}$,
and the uncertainty relation for their fluctuations is:

\begin{equation}
\Delta^{2}F_{1}\Delta^{2}F_{2}\geq\frac{1}{16},\label{eq:14}
\end{equation}
where $\Delta^{2}F_{i}=\langle\text{\ensuremath{\hat{F_{i}^{2}}}}\rangle-\langle\hat{F_{i}}\rangle^{2}$,
for $i=1,2$. Similarly to the single-mode case, two-mode squeezing
occurs if one of the variances $\triangle^{2}F_{i}$ is below the
shot noise level, i.e., $\triangle^{2}F_{i}<\frac{1}{4}$. This condition
can be satisfied if the two modes are uncorrelated, with one or both
of $F_{i}$ individually squeezed, or when nonclassical correlations,
such as entanglement between the two modes, exist. The squeezing parameter
that characterizes the quadrature squeezing of the $i$-th component
of the TMSV state is: 

\begin{equation}
Q_{i}=\Delta^{2}F_{i}-\frac{1}{4}.\label{eq:15}
\end{equation}
The values of $Q_{i}$ are bounded by $Q_{i}\geq-\frac{1}{4}$, and
the $i$-th component of the quadrature operators of the TMSV state
is said to be squeezed if $-\frac{1}{4}\leq Q_{i}<0$. After some
algebra, we obtain the quadrature squeezing parameters $Q_{i}$ for
each mode in the final state $\vert\Psi\rangle$ as follows: 

\begin{align}
Q_{1,\Psi} & =\frac{1}{4}\left[\langle\hat{a}^{\dagger}\hat{a}\rangle+\langle\hat{b}^{\dagger}\hat{b}\rangle+\Re\left[\langle\hat{a}^{2}\rangle\right]+\Re\left[\langle\hat{b}^{2}\rangle\right]\right]\nonumber \\
 & +\frac{1}{2}\left[\Re\left[\langle\hat{a}\hat{b}\rangle\right]+\Re\left[\langle\hat{a}^{\dagger}\hat{b}\rangle\right]\right]-\frac{1}{2}\left[\Re\left[\langle\hat{a}\rangle\right]+\Re\left[\langle\hat{b}\rangle\right]\right]^{2}.\label{eq:16}\\
Q_{2,\Psi} & =\frac{1}{4}\left[\langle\hat{a}^{\dagger}\hat{a}\rangle+\langle\hat{b}^{\dagger}\hat{b}\rangle-\Re\left[\langle\hat{a}^{2}\rangle\right]-\Re\left[\langle\hat{b}^{2}\rangle\right]\right]\nonumber \\
 & +\frac{1}{2}\left[\Re\left[\langle\hat{a}^{\dagger}\hat{b}\rangle\right]-\Re\left[\langle\hat{a}\hat{b}\rangle\right]\right]-\frac{1}{2}\left[\Im\left[\langle\hat{a}\rangle\right]+\Im\left[\langle\hat{b}\rangle\right]\right]^{2}.\label{eq:17}
\end{align}
Here, $\langle\cdots\rangle$ denotes the expectation values of the
associated operators under the state $\vert\Psi\rangle$, and we take
$\vartheta=0$. The Appendix \ref{sec:A1} listed the analytic expressions
for the expectation values of the operators. If we set $s=0$, the
above expressions simplify to: 
\begin{align}
Q_{1,\phi} & =\frac{1}{4}\left(e^{2\lambda}-1\right),\\
Q_{2,\phi} & =\frac{1}{4}\left(e^{-2\lambda}-1\right).
\end{align}
These squeezing parameters $Q_{i,\phi}$ correspond to the initial
state $\vert\phi\rangle$. Initially, there is no squeezing effect
along $F_{1}$, while squeezing exists along the $F_{2}$ axis, and
$Q_{2,\phi}$ reaches its maximum squeezing value of $-0.25$ as the
squeezing parameter $\lambda$ increases. Since the explicit expressions
for the quadrature squeezing parameters $Q_{1,\Psi}$ and $Q_{2,\Psi}$
are too complex for analytical treatment, we provide numerical results
in the following. Figures. \ref{fig:Q1-B-1} and \ref{fig:Q2-C} show
the numerical results of the quadrature squeezing parameters for different
system parameters.

From the results in Figs. \ref{fig:Q1-B-1}(a) and (b), we deduce
that after the postselected von Neumann measurement, a squeezing effect
occurs along the $F_{1}$ quadrature. This effect increases in the
WM regime with anomalous weak values for small squeezing parameter
values $\lambda$ ($0<\lambda\apprle0.28$). This positive result
is due to the signal amplification effect of WM.

As shown in Fig. \ref{fig:Q2-C}, the anomalously large weak values
harm $Q_{2}$, consistent with the concept that more squeezing along
one axis results in less squeezing along the other. These numerical
results indicate that stronger squeezing occurs for $Q_{1}$ as weak
values increase in the weak coupling regime, highlighting the role
of WMs in amplifying signals in this region. However, as the coupling
strength parameter $s$ increases, pushing the system gradually into
the strong measurement regime, the large weak values no longer enhance
the squeezing properties in $Q_{1}$ and $Q_{2}$. The quadrature
squeezing of the TMSV state can be experimentally verified using homodyne
detection techniques \citep{PhysRevA.82.021801}. From these numerical
results, we can confirm that the postselected von Neumann measurement,
characterized by postselection and weak values, positively affects
the enhancement of the squeezing effects of the TMSV state. However,
it is important to note that the squeezing effect is phase-dependent,
which means experimental measures of nonclassicality based solely
on squeezing parameters may sometimes fail to detect the quantumness
of a given field. Besides squeezing, there are various other criteria
to check the nonclassicality of a field, such as the Mandel-$Q$ factor
$Q_{M}$, second-order correlations, and entanglement (in the case
of multi-mode systems), among others. Thus, further investigation
of these criteria is necessary to provide a more comprehensive confirmation
of the enhancement of nonclassicality in the TMSV state when using
postselected von Neumann measurements.
\begin{center}
\begin{figure}
\begin{centering}
\includegraphics[width=8cm]{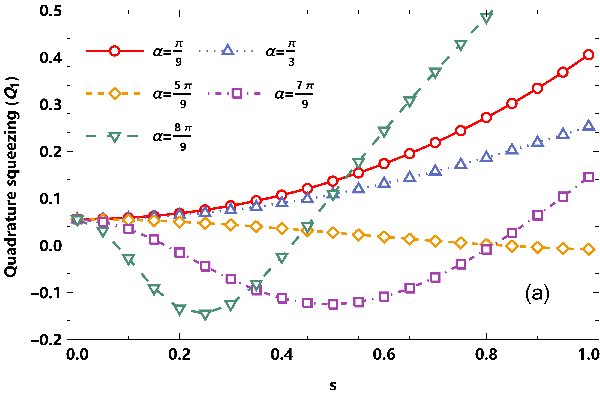}
\par\end{centering}
\begin{centering}
\includegraphics[width=8cm]{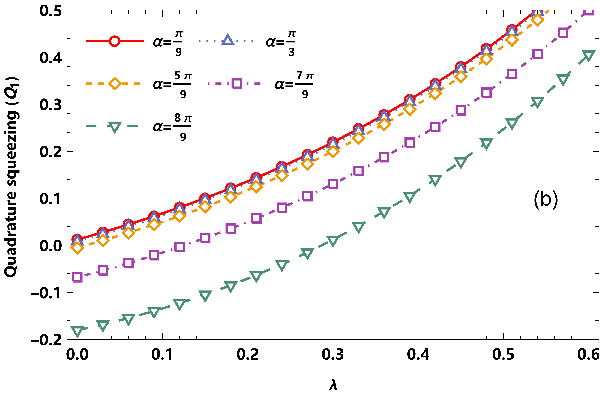}
\par\end{centering}
\caption{\label{fig:Q1-B-1}$Q_{1}$of the state $\vert\Psi\rangle$ for different
system parameters. (a) $Q_{1}$ as a function of coupling strength
parameter $s$ for different weak values while the two-mode squeezing
parameter is $\lambda=0.1$. (b) $Q_{1}$ as a function of squeezing
parameter $\lambda$ for different weak values while the coupling
strength parameter is $s=0.2$. In both cases, we take $\delta=\theta=\vartheta=0$.}
\end{figure}
\par\end{center}

\begin{center}
\begin{figure}
\begin{centering}
\includegraphics[width=8cm]{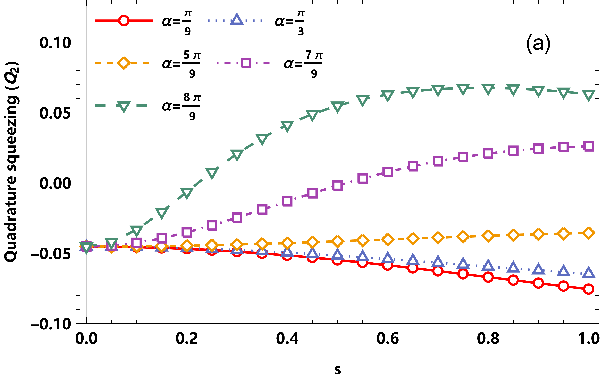}
\par\end{centering}
\begin{centering}
\includegraphics[width=8cm]{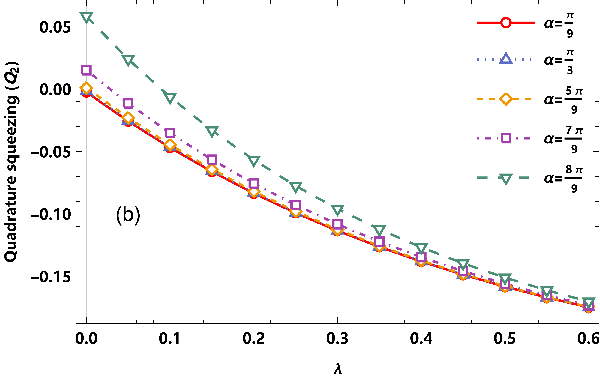}
\par\end{centering}
\caption{\label{fig:Q2-C}$Q_{2}$ of the state $\vert\Psi\rangle$ for different
system parameters. (a) $Q_{2}$ as a function of coupling strength
parameter $s$ for different weak values while the two-mode squeezing
parameter is $\lambda=0.1$. (b) $Q_{2}$ as a function of squeezing
parameter $\lambda$ for different weak values while the coupling
strength parameter is $s=0.2$. Other parameters are the same as in
Fig. \ref{fig:Q1-B-1}.}
\end{figure}
\par\end{center}

\section{The effects on quanutm statistics\label{sec:4}}

\subsection{Mandel-$Q$ parameter\label{subsec:4-1}}

Researchers often use the sub-Poissonian statistical distribution
of a light field as a criterion for identifying nonclassicality. Based
on the distribution of the Mandel-$Q$ parameter, three types of distributions
can be distinguished: sub-Poissonian, super-Poissonian, and Poissonian.
In this subsection, we primarily analyze the photon statistical distribution
of the TMSV states via the Mandel-$Q$ parameter, which is defined
as follows \citep{Agarwal2013}:

\begin{equation}
Q_{a,\Psi}=\frac{\langle\hat{a}^{\dagger2}\hat{a}^{2}\rangle-\langle\hat{a}^{\dagger}\hat{a}\rangle^{2}}{\langle\hat{a}^{\dagger}\hat{a}\rangle}.\label{eq:20-1}
\end{equation}
In principle, $Q_{a}<0$ indicates a sub-Poissonian distribution,
implying the presence of nonclassicality. Below, we focus on the condition
$Q_{a}<0$ for the generated state. We obtain the explicit expression
of $Q_{a}$ for the state $\vert\Psi\rangle$ by substituting the
relevant expectation values listed in Appendix \ref{sec:A1}. By setting
$s=0$, we find that the Mandel-$Q$ parameter for the TMSV state
$\vert\phi\rangle$ is $Q_{a,\phi}=Q_{b,\phi}=\sinh^{2}\lambda$.
This result corresponds to a super-Poissonian distribution, as $Q_{a,\phi}>0$,
except for the trivial case where $\lambda=0$. In this case, there
isn't a definitive conclusion about the nonclassicality of the state.

To clearly understand the effects of postselected von Neumann measurements
on the quantum statistics of the TMSV state, we rely on numerical
analysis, with the results shown in Fig. \ref{fig:mandel}. In Fig.
\ref{fig:mandel} (a), we fix the weak value parameter $\alpha=8\pi/9$
and plot $Q_{a,\Psi}$ as a function of the squeezing parameter $\lambda$
for different values of the coupling strength parameter $s$. As shown
in the figure, after applying postselected measurements ($s\neq0$),
$Q_{a,\Psi}$ becomes negative for squeezing parameter $\lambda$
in the range $0<\lambda\lessapprox0.7$, especially when considering
large weak values. This result indicates that postselected WMs with
anomalous weak values are beneficial for improving the quantum statistics
of the TMSV state. To further confirm this result, Fig. \ref{fig:mandel}
(b) illustrates $Q_{a,\Psi}$ as a function of the coupling strength
parameter $s$, setting $\lambda=0.1$ for different weak values characterized
by $\alpha$. We observe that $Q_{a,\Psi}<0$ for large weak values,
and its negative values increase as the coupling strength parameter
$s$ grows.

We know that the TMSV state $\vert\phi\rangle$ is a Gaussian continuous-variable
entangled state, and its Wigner function does not exhibit regions
in phase space distribution with negative values. However, with the
postselected von Neumann measurement, the quantum statistics of the
TMSV state change from super-Poissonian to sub-Poissonian, significantly
enhancing its nonclassicality. This result can also be seen clearly
through the zero-time delay second-order correlation function $g^{(2)}(0)$,
which we will discuss in detail in the next subsection.
\begin{center}
\begin{figure}
\begin{centering}
\includegraphics[width=8cm]{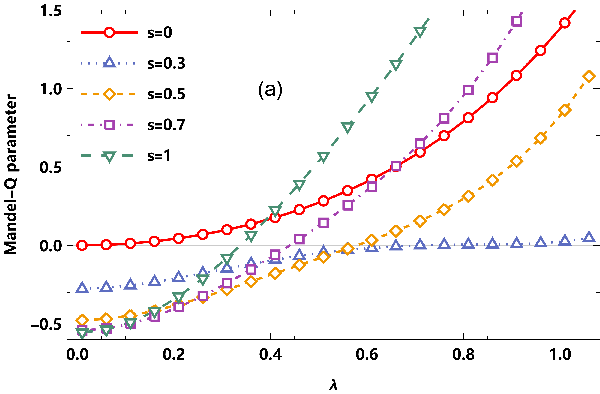}
\par\end{centering}
\begin{centering}
\includegraphics[width=8cm]{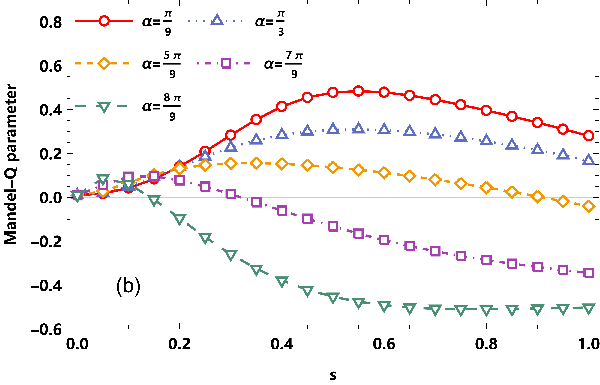}
\par\end{centering}
\caption{\label{fig:mandel}Mandel-$Q_{a,\Psi}$ parameter of TMSV state under
$\vert\Psi\rangle$ for different system parameters. (a) Mandel-$Q_{a,\Psi}$
parameter as a function of the squeezing parameter $\lambda$ for
different values of the coupling strength parameter $s$, with the
weak value parameter fixed at $\alpha=\frac{8\pi}{9}$; (b) Mandel-$Q_{a,\Psi}$
parameter as a function of coupling strength parameter $s$ for different
weak values, with the squeezing parameter fixed at $\lambda=0.1$.}
\end{figure}
\par\end{center}

\subsection{Second-order correlation function\label{subsec:4-2}}

We can also examine the statistical properties of the radiation field
using the zero-time delay second-order correlation function, $g^{(2)}(0)$.
When $0\le g^{(2)}(0)<1$, the field exhibits sub-Poissonian statistics,
which serves as an indicator of the non-classical nature of the radiation
field \citep{Agarwal2013}. For our state $\vert\Psi\rangle$, the
second-order correlation function $g^{(2)}(0)$ for its two modes
is expressed as follows: 
\begin{equation}
g_{a,\Psi}^{(2)}(0)=\frac{\langle\hat{a}^{\dagger2}\hat{a}^{2}\rangle}{\langle\hat{a}^{\dagger}\hat{a}\rangle^{2}}
\end{equation}
 and 
\begin{equation}
g_{b,\Psi}^{(2)}(0)=\frac{\langle\hat{b}^{\dagger2}\hat{b}^{2}\rangle}{\langle\hat{b}^{\dagger}\hat{b}\rangle^{2}}.
\end{equation}
We can observe that the condition $g_{a,\Psi}^{(2)}(0)<1$ is equivalent
to the negativity of the $Q$-parameter $Q_{a,\Psi}$ in Eq. (\ref{eq:20-1}).
Indeed, the $Q$-parameter and $g^{(2)}(0)$ are related by the expression
$Q=\langle\hat{n}\rangle\left[g^{(2)}(0)-1\right]$. For the TMSV
state $\vert\phi\rangle$, this relationship yields $g_{a,\phi}^{(2)}(0)$=$g_{b,\phi}^{(2)}(0)=2$.
We can obtain the explicit form of $g^{(2)}(0)$ by calculating the
expectation values of the relevant operators, which we listed in Appendix
\ref{sec:A1}. Fig. \ref{fig:g2(0)} shows the behavior of $g^{(2)}(0)$
as a function of the squeezing parameter $\lambda$ for different
values of the coupling strength parameter $s$. To explore the effects
of large anomalous weak values on the statistical properties of the
state, we fix the weak value parameter to $\alpha=8\,\pi/9\,$ in
these plots. As shown in Fig. \ref{fig:g2(0)}, the case $s=0$ corresponds
to the initial TMSV state's $g^{(2)}(0)$, which remains constant
with respect to $\lambda$. This constant value equals the second-order
correlation function of a thermal state, $g_{th}^{(2)}(0)=2$. We
can explain this phenomenon: the reduced density matrix of one mode
of the TMSV state, obtained by tracing out the other mode, results
in a thermal state
\begin{equation}
\rho_{th}=\left(1-\eta\right)\sum_{n=0}^{\infty}\eta^{n}\vert n\rangle\langle n\vert=\sum_{n=0}^{\infty}\frac{\langle n\rangle_{th}^{n}}{\left(1+\langle n\rangle_{th}\right)^{n+1}}\vert n\rangle\langle n\vert.\label{eq:24}
\end{equation}
Here, $\eta=\tanh^{2}\lambda$ and $\langle\hat{n}\rangle_{th}=Tr\left(\hat{a}^{\dagger}\hat{a}\rho_{th}\right)=Tr\left(\hat{b}^{\dagger}\hat{b}\rho_{th}\right)$
represents the average number of excitations. 

We can show that the second-order correlation function at zero-time
delay, $g^{(2)}(0)$, for a thermal state is exactly equal to two.
From Fig. \ref{fig:g2(0)}, we can observe that when there is an interaction
between the measured system and the measuring device ($s\neq0$),
the correlation functions $g^{(2)}(0)$ of the different modes of
$\vert\Psi\rangle$ exhibit distinct characteristics. For large values
of the squeezing parameter ($\lambda\gg1$), $g_{a}^{(2)}(0)$ and
$g^{(2)}(0)$ become indistinguishable and tend toward the value corresponding
to a thermal state. In contrast, for small squeezing parameters, the
two modes of $\vert\Psi\rangle$ present opposite statistics: while
mode $a$ exhibits antibunching $(g_{a}^{(2)}(0)<1)$, indicating
sub-Poissonian statistics, mode $b$ shows superbunching $(g_{b}^{(2)}(0)>2)$.
Specifically, $g_{a}^{(2)}(0)<1$ for squeezing parameter $\lambda\lessapprox0.7$,
and $g_{b}^{(2)}(0)>2$ for $\lambda\lessapprox0.8$. Therefore, we
can conclude that postselected von Neumann measurement significantly
alters the photon statistics for large weak values.

Since bosonic superbunching has potential applications in technical
areas such as ghost interference, imaging, and efficient nonlinear
light-matter interactions \citep{BD,17,PhysRevA.95.053809,LIU2018824,PhysRevA.97.053835,PhysRevX.8.011013,PhysRevA.99.063827},
the $b$ mode of our generated state $\vert\Psi\rangle$ is particularly
well-suited for quantum devices related to advanced imaging techniques.
On the other hand, the $a$ mode of $\vert\Psi\rangle$ may be useful
for single-photon generation \citep{RevModPhys.54.1061,Lounis_2005}. 

In the previous two subsections, we investigated the quantum statistics
of our output state $\vert\Psi\rangle$ through the Mandel-$Q$ parameter
and the zero-time delay second-order correlations for each individual
mode. However, the two modes of the TMSV state are not independent
entities, as they inherently possess non-zero quantum correlations
due to $\langle\hat{a}\hat{b}\rangle=\frac{1}{2}\sinh(2\lambda)e^{i\theta}$.
Therefore, it is essential to explore the effects of postselected
von Neumann measurement on the quantum correlations of the TMSV state
by considering both modes simultaneously. 
\begin{center}
\begin{figure}
\begin{centering}
\includegraphics[width=8cm]{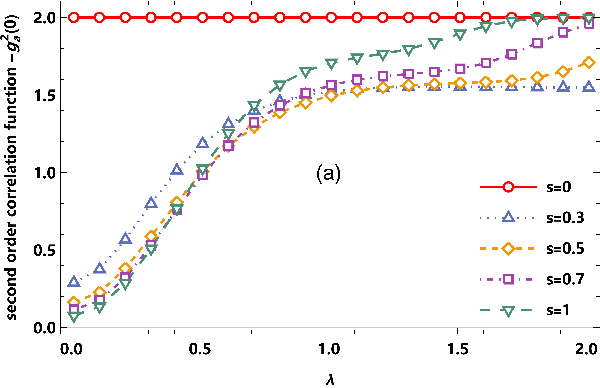}
\par\end{centering}
\begin{centering}
\includegraphics[width=8cm]{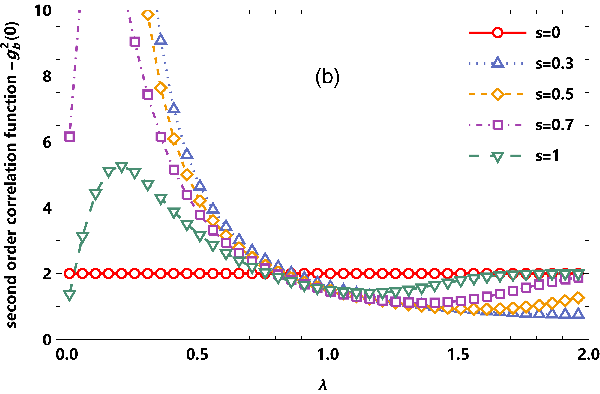}
\par\end{centering}
\caption{\label{fig:g2(0)} Second-order correlation function $g^{(2)}(0)$
of the state $\Psi$ for different system parameters. (a) $g_{a}^{(2)}(0)$
and (b) $g_{b}^{(2)}(0)$ as a function of squeezing parameter $\lambda$
for different coupling strength parameter $s$ while fixed the weak
value parameter $\alpha=\frac{8\pi}{9}$. Other parameters are the
same as in Fig. \ref{fig:Q1-B-1}.}
\end{figure}
\par\end{center}

\subsection{Second-order cross-correlation function\label{subsec:4-3}}

In this subsection, we study the second-order cross-correlation (SOCC)
function $g_{a,b}^{(2)}$ of the TMSV state after postselected von
Nuemann measurement by using the $\vert\Psi\rangle$ state. The SOCC
function of two-mode radiation field is defined as \citep{REN2019106,GIRI2017140} 

\begin{equation}
g_{a,b,\Psi}^{(2)}=\frac{\langle\hat{a}^{\dagger}\hat{a}\hat{b}^{\dagger}\hat{b}\rangle}{\langle\hat{a}^{\dagger}\hat{a}\rangle\langle\hat{b}^{\dagger}\hat{b}\rangle}.\label{eq:20}
\end{equation}
 Here, $\langle\hat{a}^{\dagger}\hat{a}\hat{b}^{\dagger}\hat{b}\rangle$
represents the intensity-intensity correlation between the two-modes,
while $\langle\hat{a}^{\dagger}\hat{a}\rangle$ and $\langle\hat{b}^{\dagger}\hat{b}\rangle$
are denote mean photon number of photons for each mode, respectively.
This function characterizes the correlation between photons in the
different modes. If $g_{a,b,\Psi}^{(2)}>1$, there exists correlation
between the $a$-mode and $b$-mode of the two-mode radiation field.
Otherwise, they are inversely correlated.\textcolor{blue}{{} }To investigate
the properties of $g_{a,b,\Psi}^{(2)}$, we first derive the average
values of $\langle\hat{a}^{\dagger}\,\hat{a}\,\hat{b}^{\dagger}\,\hat{b}\rangle$,
$\langle\hat{a}^{\dagger}\,\hat{a}\rangle$, and $\langle\hat{b}^{\dagger}\,\hat{b}\rangle$
under the state $\vert\Psi\rangle$. Since their explicit expressions
are cumbersome, we list them in Appendix \ref{sec:A1}. Specifically,
when $s=0$, the function $g_{a,b,\Psi}^{(2)}$ reduces to the SOCC
function of the initial TMSV state $\vert\phi\rangle$, and is expressed
as

\begin{align}
g_{a,b,\phi}^{(2)} & =\frac{\sinh^{2}\lambda\cosh^{2}\lambda+\sinh^{4}\lambda}{\sinh^{4}\lambda}=\coth^{2}\lambda+1.\label{eq:26}
\end{align}
Eq. \ref{eq:26} shows that the TMSV state always has strong correlations
between the two modes, particularly in the small $\lambda$ regime.
It is also clear that as the squeezing parameter $\lambda$ increases,
$g_{a,b,\phi}^{(2)}$ approaches 2. To further examine the effect
of postselected von Neumann measurements on the SOCC of the TMSV state,
we plot $g_{a,b,\Psi}^{(2)}$ for different system parameters associated
with the postselected von Neumann measurement, with corresponding
numerical results shown in Fig. \ref{fig:G2-1}. We plot $g_{a,b,\Psi}^{(2)}$
as a function of the two-mode squeezing parameter $\lambda$ (Fig.
\ref{fig:G2-1}(a)) and the coupling strength parameter $s$ (Fig.
\ref{fig:G2-1}(b)), respectively.

In the WM regime ($s\ll1$), large anomalous weak values positively
affect the enhancement of SOCC for the TMSV state. Interestingly,
we observe that $g_{a,b,\Psi}^{(2)}$ always tends to the fixed value
of 2 for large values of the squeezing parameter $\lambda$, regardless
of the weak value $\langle\hat{\sigma}_{x}\rangle_{w}$ or the coupling
strength parameter $s$ (see Fig. \ref{fig:G2-1}(a)). This result
shows that the photons in the two different modes maintain constant
correlation even after the postselected von Neumann measurement.

We can explain this trend as follows: For the TMSV state, the correlation
function $\langle\hat{a}^{\dagger}\hat{a}\hat{b}^{\dagger}\hat{b}\rangle$
is well-known and can be written as \citep{PhysRevResearch.3.033095}
\begin{equation}
\langle\hat{a}^{\dagger}\hat{a}\hat{b}^{\dagger}\hat{b}\rangle=\langle\hat{a}^{\dagger}\hat{a}\rangle\langle\hat{b}^{\dagger}\hat{b}\rangle+\vert\langle\hat{a}\hat{b}\rangle\vert^{2},\label{eq:22}
\end{equation}
and the correlation can be further analyzed using the Cauchy-Schwartz
inequality (CSI): 
\begin{equation}
\vert\langle\hat{a}\hat{b}\rangle\vert^{2}\le\langle\hat{a}^{\dagger}\hat{a}\rangle\langle\hat{b}\hat{b}^{\dagger}\rangle=\langle\hat{a}^{\dagger}\hat{a}\rangle\left(\langle\hat{b}^{\dagger}\hat{b}\rangle+1\right)\label{eq:23}
\end{equation}
By considering Eq. (\ref{eq:22}) and Eq. (\ref{eq:23}), we can rewrite
the SOCC function $g_{a,b,\Psi}^{(2)}$, defined in Eq. (\ref{eq:20}),
as
\begin{equation}
g_{a,b,\Psi}^{(2)}=\frac{\langle\hat{a}^{\dagger}\hat{a}\hat{b}^{\dagger}\hat{b}\rangle}{\langle\hat{a}^{\dagger}\hat{a}\rangle\langle\hat{b}^{\dagger}\hat{b}\rangle}\le2+\frac{1}{\langle\hat{b}^{\dagger}\hat{b}\rangle}.\label{eq:24-1}
\end{equation}
The explicit expression for $\langle\hat{b}^{\dagger}\hat{b}\rangle$
is provided in Appendix \ref{sec:A1}, and it increases with the squeezing
parameter $\lambda$ such that $\langle\hat{b}^{\dagger}\hat{b}\rangle\gg1$.
In this case, $g_{a,b,\Psi}^{(2)}$ reaches its minimum value of 2.

Moreover, as we focus on smaller values of $\lambda$, the SOCC function
can exceed one for small coupling strength parameters $s$ and large
weak values, indicating strong correlations. Another interesting observation
is that when $g_{a,b,\Psi}^{(2)}$ is plotted as a function of the
coupling strength parameter $s$ (see Fig. \ref{fig:G2-1}(b)) for
a fixed small squeezing parameter $\lambda$ (specifically, $\lambda=0.1$),
the SOCC function decreases with increasing $s$ and approaches 1.

From the expression for the state $\vert\Psi\rangle$ given in Eq.
(\ref{eq:Psi}), we can deduce that when the squeezing parameter $\lambda$
is too small, the effect of the two-mode squeezing operator $\hat{S}(\xi)$
on the state $\vert\Psi\rangle$ becomes trivial. Meanwhile, the displacement
operator $D\left(\pm\frac{s}{2}\right)$ becomes dominant as the coupling
strength parameter $s$ increases. Since the squeezing parameter $\lambda$
is small and the displacement operator only acts on the $a$ mode
of the TMSV state, the correlation $\langle\hat{a}\,\hat{b}\rangle$
between modes $a$ and $b$ approximately approaches zero. Consequently,
we have $\langle\hat{a}^{\dagger}\,\hat{a}\,\hat{b}^{\dagger}\,\hat{b}\rangle\approx\langle\hat{a}^{\dagger}\,\hat{a}\rangle\langle\hat{b}^{\dagger}\,\hat{b}\rangle\quad\text{[see Eq. (\ref{eq:22})]}.$
Thus, the value of $g_{a,b,\Psi}^{(2)}$ equals one in this extreme
case.

From the numerical results discussed above, we can confirm that after
the postselected von Neumann measurement, the quantum correlation
between the two modes of our output state $\vert\Psi\rangle$ is significantly
enhanced, especially in WM regimes. We can further verify this result
by examining various entanglement criteria. 
\begin{center}
\begin{figure}
\begin{centering}
\includegraphics[width=8cm]{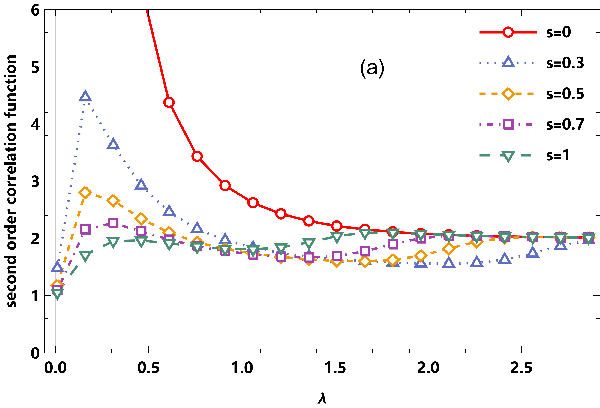}
\par\end{centering}
\begin{centering}
\includegraphics[width=8cm]{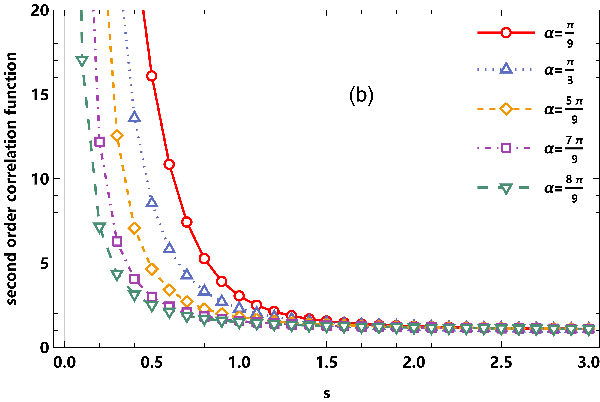}
\par\end{centering}
\caption{\label{fig:G2-1}SOCC function of the TMSV state under $|\Psi\rangle$
for different system parameters. (a) SOCC function as a function of
squeezing parameter $\lambda$ for different coupling strength parameters
$s$ while fixing the weak value parameter $\alpha=\frac{8\pi}{9}$;
(b) SOCC function as a function of the coupling strength parameter
$s$ for different weak values while keeping the two-mode squeezing
parameter $\lambda=0.1$. Other parameters are the same as in Fig.
\ref{fig:Q1-B-1}.}
\end{figure}
\par\end{center}

\section{The effects on entanglement\label{sec:5}}

In this section, we further investigate the effects of postselected
von Neumann measurement on the quantum correlations of the TMSV state.
Specifically, we examine the entanglement between the two modes and
the Einstein-Podolsky-Rosen (EPR) correlations of the TMSV state after
measurement, considering observables quadratic in the creation and
annihilation operators.

\subsection{Cauchy--Schwarz inequality\label{subsec:5-1}}

The Cauchy-Schwarz inequalities have played a significant role in
quantum mechanics. In quantum optics, we determine the nonclassicality
of radiation fields by the violation of the Cauchy-Schwarz inequality
(CSI). For a two-mode light field, the CSI is \citep{Buzek1992TheOO,Agarwal2013}

\begin{equation}
\left\langle \hat{a}^{\dagger2}\hat{a}^{2}\right\rangle \left\langle \hat{b}^{\dagger2}\hat{b}^{2}\right\rangle \geqslant\left|\left\langle \hat{a}^{\dagger}\hat{a}\hat{b}^{\dagger}\hat{b}\right\rangle ^{2}\right|.
\end{equation}
In analogy to the Mandel-$Q$ parameter, we can introduce a parameter
$I_{0}$ defined as

\begin{equation}
I_{0}=\frac{\left(\left\langle \hat{a}^{\dagger2}\hat{a}^{2}\right\rangle \left\langle \hat{b}^{\dagger2}\hat{b}^{2}\right\rangle \right)^{1/2}}{\left|\left\langle \hat{a}^{\dagger}\hat{a}\hat{b}^{\dagger}\hat{b}\right\rangle \right|}-1.
\end{equation}
 The nonclassicality of the two-mode state can be assessed using $I_{0}$.
Notably, $I_{0}=0$ for coherent states, $I_{0}>0$ for classical
states, while $-1\leq I_{0}<0$ indicates purely nonclassical behavior.
Thus, negative values of $I_{0}$ imply the nonclassicality of the
corresponding two-mode radiation field. For our analysis, we can derive
the expression of $I_{0,\Psi}$ for the state $\vert\Psi\rangle$
using the average values of $\langle\hat{a}\,\hat{b}\rangle$, $\langle\hat{a}^{\dagger2}\,\hat{a}^{2}\rangle$,
and $\langle\hat{b}^{\dagger2}\,\hat{b}^{2}\rangle$ listed in Appendix
\ref{sec:A1}.

To investigate the effects of postselected von Neumann measurement
on the parameter $I_{0}$, we plot $I_{0,\Psi}$ for various system
parameters, and the numerical results are presented in Fig. \ref{fig:4}.
When $s=0$, the above expression simplifies to the CSI parameter
$I_{0,\phi}$ for the initial state $\vert\phi\rangle$:
\begin{equation}
I_{0,\phi}=\frac{\sinh^{2}\lambda}{2\sinh^{2}\lambda+1}-1.\label{eq:38}
\end{equation}
 As shown by the red solid circle curve in Fig. \ref{fig:4} (a),
the $I_{0,\phi}$ of the initial state $\vert\phi\rangle$ exhibits
very strong nonclassical features when $\lambda\ll1$. As $\lambda$
increases, its nonclassicality gradually weakens and approaches semiclassical
behavior. The Fig. \ref{fig:4} illustrates that the postselected
von Neumann measurement positively affects the nonclassicality of
the TMSV state. In contrast to the initial state, the output state
$\vert\Psi\rangle$ after postselected von Nuemann measurement retains
significant nonclassicality even when $\lambda>1$ in the WM regime
with anomalous weak values (see Fig. \ref{fig:4} (b)). Notably, as
the squeezing parameter $\lambda$ exceeds 1, the parameter $I_{0,\Psi}$
of the two-mode state $\vert\Psi\rangle$ becomes strongest at a small
coupling strength parameter $s=0.2$. Larger weak values enhance (or
diminish) nonclassicality in the two-mode state for $\lambda>1$ (or
$\lambda<1$), before transforming into a coherent state, as depicted
in Fig. \ref{fig:4} (b). This observation indicates that, in the
weak coupling regime, the quantum nature of the two-mode state after
postselected measurement becomes more evident with larger weak values
(see Figs. \ref{fig:4} (b), (c), and (d)).
\begin{center}
\begin{figure}
\begin{centering}
\includegraphics[width=8cm]{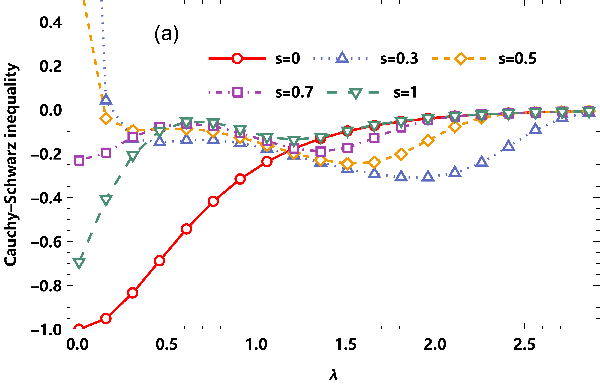}
\par\end{centering}
\begin{centering}
\includegraphics[width=8cm]{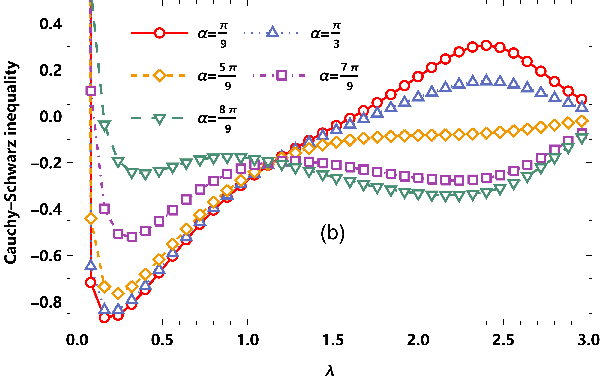}
\par\end{centering}
\begin{centering}
\includegraphics[width=8cm]{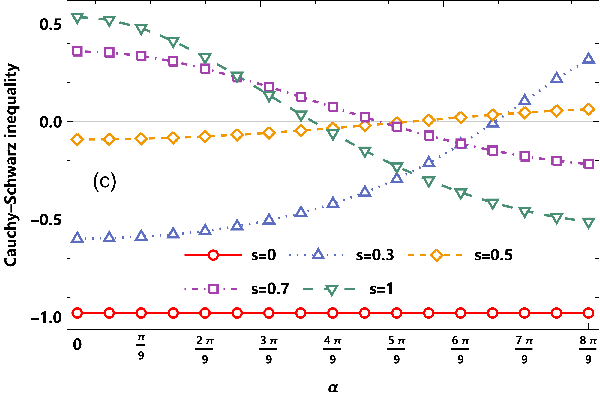}
\par\end{centering}
\begin{centering}
\includegraphics[width=8cm]{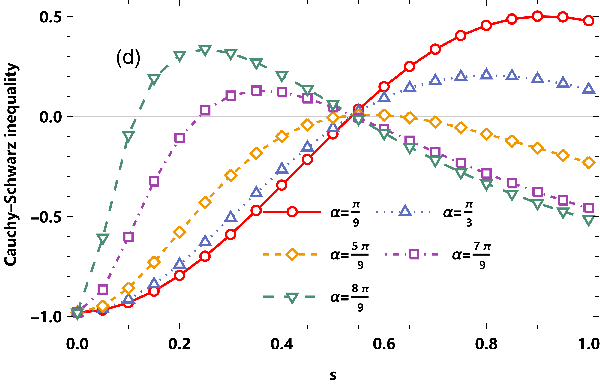}
\par\end{centering}
\caption{\label{fig:4}The parameter $I_{0,\Psi}$ of the TMSV state under
$\vert\Psi\rangle$ for different system parameters. (a) $I_{0,\Psi}$
as a function of squeezing parameter $\lambda$ for various coupling
strength parameters $s$, while the weak value parameter has a fixed
value of $\alpha=\frac{8\pi}{9}$; (b) $I_{0,\Psi}$ as a function
of squeezing parameter $\lambda$ for different weak values while
the coupling strength parameter has a fixed value of $s=0.2$; (c)
$I_{0,\Psi}$ as a function of coupling strength parameters $s$ for
various weak values while the two-mode squeezing parameter has a fixed
value of $\lambda=0.1$; (d) $I_{0,\Psi}$ as a function of weak values
characterized by $\alpha$ for different values of $s$, while keeping
$\lambda=0.1$. Other parameters are the same as in Fig.\ref{fig:Q1-B-1}.}

\end{figure}
\par\end{center}

\subsection{Hillery-Zubairy correlation\label{subsec:5-2} }

The entanglement between the two modes of the light field can be studied
using the entanglement measure proposed by Hillery and Zubairy \citep{PhysRevLett.96.050503}.
The Hillery-Zubairy correlation (HZC) characterizes the entanglement
by \citep{PhysRevA.74.032333} 

\begin{align}
E & =\langle\hat{a}^{\dagger}\hat{a}\rangle\langle\hat{b}^{\dagger}\hat{b}\rangle-\left|\langle\hat{a}\hat{b}\rangle\right|^{2}.\label{eq:entanglement-Eq}
\end{align}
This entanglement criterion highlights the importance of the correlation
$\langle\hat{a}\hat{b}\rangle$ between the two modes of a given state.
Entanglement exists between the two modes if $E<0$. Additionally,
from the CSI, we have $\vert\langle\hat{a}\hat{b}\rangle\vert^{2}\le\langle\hat{a}^{\dagger}\hat{a}\rangle\langle\hat{b}\hat{b}^{\dagger}\rangle$
\citep{PhysRevResearch.3.033095}, which leads to the conclusion that
the entanglement condition for two-mode fields is bounded by $-\langle a^{\dagger}a\rangle\le E<0.$
For the initial pointer state $\vert\phi\rangle$ the HZC criterion
is given by $E_{\phi}=-\sinh^{2}\lambda=-\langle a^{\dagger}a\rangle_{\phi}$.
This result indicates that the TMSV state $|\phi\rangle$ is a higher-order
entangled state as long as $\lambda>0$. Thus, this condition can
always detect entanglement in the state $|\phi\rangle$. 

To evaluate the HZC criterion for our output state $|\Psi\rangle$,
we calculate $E_{\Psi}$ using Eq. (\ref{eq:entanglement-Eq}), with
its explicit expression obtained by substituting the average values
of $\langle\hat{a}\hat{b}\rangle,\langle\hat{a}^{\dagger}\hat{a}\rangle$
and $\langle\hat{b}^{\dagger}\hat{b}\rangle$ listed in Appendix \ref{sec:A1}.
To clearly illustrate how the postselected von Neumann measurement
affects the entanglement of the output state $|\Psi\rangle$, we present
numerical results by plotting $E_{\Psi}$ for different parameters
of our system, as shown in Fig. \ref{fig:5}.

First, as depicted in Fig. \ref{fig:5} (a), the entanglement in the
bimodal state $|\Psi\rangle$ increases with the squeezing parameter
$\lambda$ for all $s$ values and large weak value $\langle\hat{\sigma}_{x}\rangle_{w}=5.671$,
corresponding to $\alpha=\frac{8\pi}{9}$. In Fig. \ref{fig:5} (b),
we plot $E_{\Psi}$ as a function of the squeezing parameter $\lambda$
for various weak values characterized by $\alpha$ while keeping the
coupling strength parameter $s$ in the WM regime, specifically $s=0.2$.
In this regime, the entanglement between the two bosonic modes of
the output state $|\Psi\rangle$ dramatically enhances after the postselected
von Neumann measurement compared to the initial state case for large
weak values.

To illustrate this point more clearly, in Figs. \ref{fig:5} (c) and
\ref{fig:5} (d), we plot $E_{\Psi}$ as a function of the coupling
strength parameter $s$ and the weak value angle $\alpha$, respectively.
The entanglement behaves similarly across different $s$ and $\alpha$
cases, reinforcing the fact that appropriate coupling strength parameters
$s$ and large anomalous weak values lead to more profound quantum
effects compared to the initial state $|\phi\rangle$.
\begin{center}
\begin{figure}
\begin{centering}
\includegraphics[width=8cm]{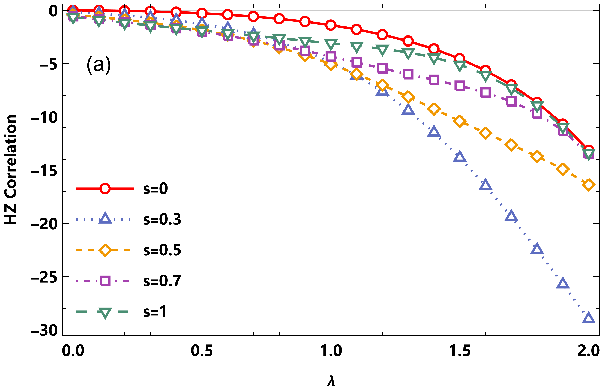}
\par\end{centering}
\begin{centering}
\includegraphics[width=8cm]{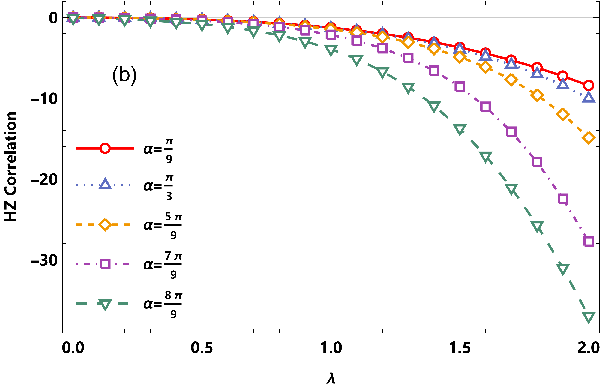}
\par\end{centering}
\begin{centering}
\includegraphics[width=8cm]{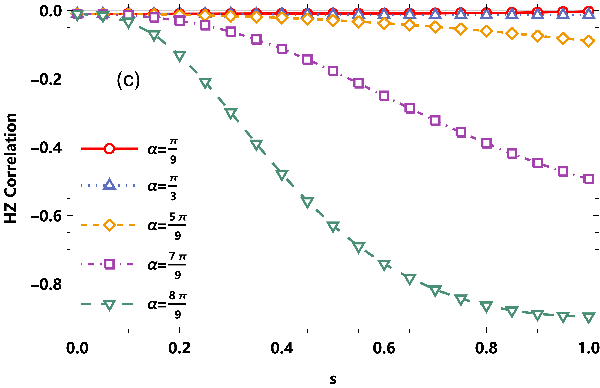}
\par\end{centering}
\begin{centering}
\includegraphics[width=8cm]{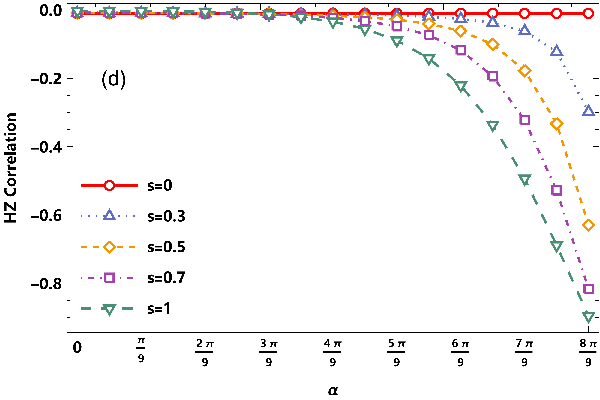}\caption{\label{fig:5}HZC function $E_{\Psi}$ of the output state $\vert\Psi\rangle$
for different system parameters. (a) $E_{\Psi}$ as a function of
squeezing parameter $\lambda$ for different coupling strength parameters
$s$, while fixing the weak value parameter $\alpha=\frac{8\pi}{9}$;
(b) $E_{\Psi}$ as a function of squeezing parameter $\lambda$ for
different weak values while fixing the coupling strength parameter
$s=0.2$; (c) $E_{\Psi}$ as a function of coupling strength parameter
$s$ for different weak values while fixing the squeezing parameter
$\lambda=0.1$; (d) $E_{\Psi}$ as a function of weak values characterized
by $\alpha$ for different values of $s$, while keeping $\lambda=0.1$.
Other parameters are the same as Fig.\ref{fig:Q1-B-1}.}
\par\end{centering}
\end{figure}
\par\end{center}

\subsection{Einstein--Podolsky--Rosen correlation\label{subsec:5-3}}

Researchers propose an inseparability criterion based on the total
variance of a pair of Einstein-Podolsky-Rosen (EPR) type operators
for continuous variable systems \citep{PhysRevLett.84.2722}. This
criterion provides a sufficient condition for entanglement of any
two-party continuous variable state, and it turns out to be a necessary
and sufficient condition for inseparability. By employing this entanglement
criterion, we calculate the total variance of EPR-type operators for
the TMSV state under the state $|\Psi\rangle$ to examine the effects
of WM (WM) on the inseparability of our system state. For the two-mode
system, the EPR correlation in terms of the variances of the EPR-type
operators $\hat{X}_{1}-\hat{X}_{2}$ and $\hat{P}_{1}+\hat{P}_{2}$
is \citep{REN2019106}

\begin{eqnarray}
\text{} & I= & \left\langle \Delta^{2}\left(\hat{X}_{1}-\hat{X}_{2}\right)\right\rangle +\left\langle \Delta^{2}\left(\hat{P}_{1}+\hat{P}_{2}\right)\right\rangle \nonumber \\
 & = & 2\left(1+\left\langle \hat{a}^{\dagger}\hat{a}\right\rangle +\left\langle \hat{b}^{\dagger}\hat{b}\right\rangle -\left\langle \hat{a}^{\dagger}\hat{b}^{\dagger}\right\rangle -\left\langle \hat{a}\hat{b}\right\rangle \right)\nonumber \\
 &  & -2\left(\left\langle \hat{a}\right\rangle -\left\langle \hat{b}^{\dagger}\right\rangle \right)\left(\left\langle \hat{a}^{\dagger}\right\rangle -\left\langle \hat{b}\right\rangle \right).\label{eq:ERP}
\end{eqnarray}
where $\hat{X}_{1}=\frac{\hat{a}+\hat{a}^{\dagger}}{\sqrt{2}},\;\hat{X}_{2}=\frac{\hat{b}+\hat{b}^{\dagger}}{\sqrt{2}}$,
$\hat{P}_{1}=\frac{\hat{a}-\hat{a}^{\dagger}}{\sqrt{2}}$ and $\text{\ensuremath{\hat{P}_{2}}}=\frac{\hat{b}-\hat{b}^{\dagger}}{\sqrt{2}}$.
If the total variance is less than 2, i.e., $0<I<2$, then one can
consider the two-mode system inseparable; otherwise, it is classical.

In order to discuss the effects of postselected von Neumann measurement
on EPR correlation, we first derive average values of associated operators
under the state $\vert\Psi\rangle$ and their values are listed in
Appendix \ref{sec:A1}. We can obtain the explicit expression of the
EPR correlation function for the state $|\Psi\rangle$ by substituting
the associated expectation values. If there is no coupling between
the measuring device and the measured system ($s=0$), then it is
easy to see that $\langle a\rangle=\langle b\rangle=0$, $\langle a^{\dagger}a\rangle=\langle b^{\dagger}b\rangle=\sinh^{2}\lambda$,
and $\langle ab\rangle=\frac{1}{2}\sinh(2\lambda)e^{i\theta}$. In
this case, the EPR correlation function reduces to $I_{s=0}=2e^{-2\lambda}\le2$,
corresponding to the TMSV state $|\phi\rangle$.

To clearly understand the effect of anomalous weak values on the EPR
correlation of $|\Psi\rangle$, we perform numerical calculations
for our discussion. In Fig. \ref{fig:6}, we plot the EPR correlation
$I$ as a function of different system parameters. In Fig. \ref{fig:6}
(a), we show the changes in $I$ concerning the squeezing parameter
$\lambda$ for different coupling strength parameters $s$ while setting
the weak value parameter $\alpha=8\pi/9$. When there is coupling
between the measuring device and the measured system ($s\neq0$),
the EPR correlation $I$ of the state $|\Psi\rangle$ is always lower
than that of $|\phi\rangle$ for anomalous weak values. To illustrate
this point more clearly, we plot the EPR correlation function $I$
as a function of the squeezing parameter $\lambda$ and the coupling
strength parameter $s$ in Figs.\ref{fig:6} (b) and \ref{fig:6}
(c), respectively, for different weak values characterized by $\alpha$.
The numerical results presented in Figs.\ref{fig:6} (b) and \ref{fig:6}
(c) indicate that the value of $I$ decreases with increasing weak
values across all regions of the squeezing parameter $\lambda$ and
the coupling strength parameter $s$, respectively. We also confirm
that large weak values positively affect the enhancement of EPR correlation,
even if we only apply postselected WMs on one mode of the TMSV.

In summary, from Figs. \ref{fig:6} (a)-\ref{fig:6} (c), it is clear
that after postselected von Neumann measurement, the inseparability
between the two modes of the state $|\Psi\rangle$ is enhanced compared
to the initial TMSV state $|\phi\rangle$, as long as the squeezing
parameter $\lambda$ is nonzero and the weak value is not very small.
This result implies that after postselected von Neumann measurement,
all parameter regions satisfy the inequality $2e^{-2\lambda}-I>0$,
except for small $\alpha$ and $\lambda\neq0$.
\begin{center}
\begin{figure}
\begin{centering}
\includegraphics[width=8cm]{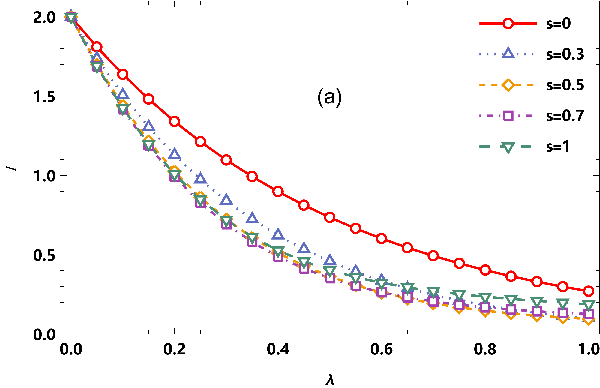}
\par\end{centering}
\begin{centering}
\includegraphics[width=8cm]{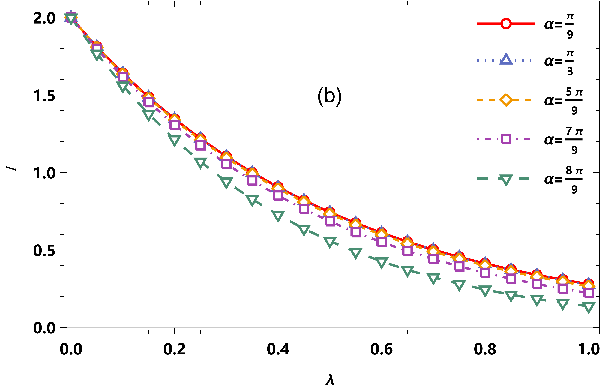}
\par\end{centering}
\begin{centering}
\includegraphics[width=8cm]{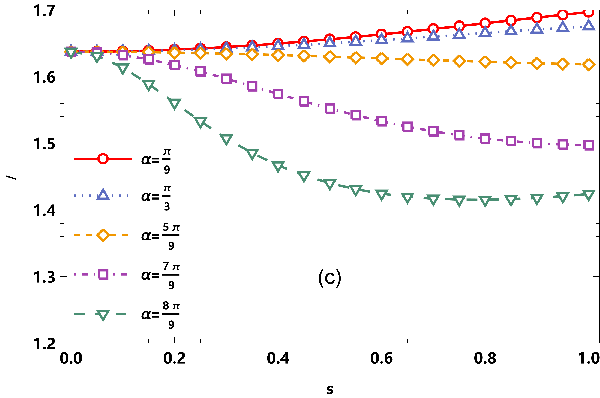}
\par\end{centering}
\caption{\label{fig:6}EPR correlation of the TMSV state under $\vert\Psi\rangle$
for different system parameters. (a) EPR correlation as a function
of squeezing parameter $\lambda$ for different weak values while
fixing the coupling strength parameter $s=0.2$; (b) EPR correlation
as a function of coupling strength parameter $s$ for different weak
values while fixing the two-mode squeezing parameter $\lambda=0.1$;
(c) EPR correlation as a function of weak values characterized by
$\alpha$ for different $s$ while keeping $\lambda=0.1$. Other parameters
are the same as in Fig.\ref{fig:Q1-B-1}.}
\end{figure}
\par\end{center}

\section{Fidelity\label{sec:6}}

In this section, we examine the state distance between the initial
and final pointer states, namely $\vert\phi\rangle$ and $\vert\Psi\rangle$,
in terms of the fidelity function. We define the fidelity $F$ as
the square of the absolute value of their scalar product, and its
value has bounds: $0\leq F\leq1$, where the lower (higher) boundary
represents different (the same) states. This quantity is a natural
candidate for the state distance because it corresponds to the similarity
of states in the Hilbert space. We calculate the fidelity as 

\begin{eqnarray}
F & = & \vert\langle\phi\vert\Psi\rangle\vert^{2}=\vert\kappa\vert^{2}\exp\left[-\frac{s^{2}\cosh(2\lambda)}{2}\right],\label{eq:42}
\end{eqnarray}
where $\kappa$ is the normalization coefficient defined in Sec. II.
This expression shows that the fidelity $F$ decreases exponentially
with increasing coupling strength parameter $s$ and squeezing parameter
$\lambda$. We plot the fidelity function $F$ for different system
parameters to illustrate how the initial TMSV state $\vert\phi\rangle$
changes via postselected von Neumann measurement; Fig. \ref{fig:7}
presents the numerical results.

As shown in Figs. \ref{fig:7} (a) and (b), the postselected von Neumann
measurement indeed leads to a change in the given state, and the state
distance between $\vert\phi\rangle$ and $\vert\Psi\rangle$ becomes
larger, even changing to a distinguishable state with increasing coupling
strength parameter $s$ and squeezing parameter $\lambda$, regardless
of how large or small the weak values are. However, larger analogous
weak values have a more positive effect than smaller ones for distinctly
transforming between states.
\begin{center}
\begin{figure}
\begin{centering}
\includegraphics[width=8cm]{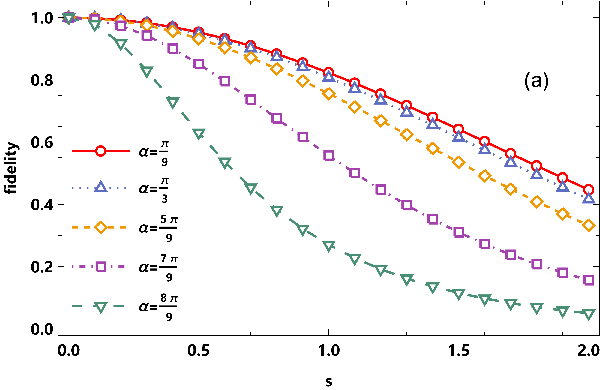}
\par\end{centering}
\begin{centering}
\includegraphics[width=8cm]{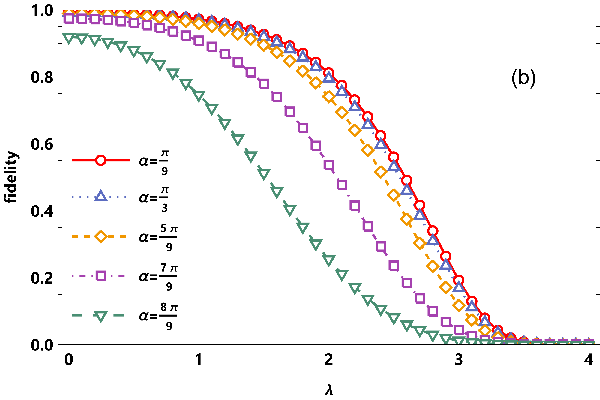}
\par\end{centering}
\caption{\label{fig:7}(a) Fidelity function between the initial and final
pointer states as a function of coupling strength parameter $s$ for
different weak values while fixing the two-mode squeezing parameter
$\lambda=0.1$; (b) Fidelity function as a function of squeezing parameter
$\lambda$ for different weak values while fixing the coupling strength
parameter $s=0.2$.}
\end{figure}
\par\end{center}

The coupling strength parameter $s$ varies continuously, and as mentioned
in Sec. II, it characterizes the transition from weak to strong measurements.
We refer to the measurement as a postselected weak (strong) measurement
when $s\ll1$ ($s\gg1$). Specifically, as $s\rightarrow0$ ($s\rightarrow\infty$),
the fidelity function approaches $F\rightarrow1$ ( $F\rightarrow0$),
indicating a transition from the postselected WM regime to the postselected
strong measurement regime. Therefore, a comprehensive discussion of
the fidelity function $F$ across weak and strong measurement regimes
provides deeper insights into our theoretical results.

When considering the coupling strength parameter $s$ to be sufficiently
small, i.e., $s\ll1$, the fidelity function in Eq. (\ref{eq:42})
reduces to
\begin{equation}
F\approx\frac{4-2s^{2}\cosh(2\lambda)}{4+s^{2}\cosh(2\lambda)\vert\langle\hat{\sigma}_{x}\rangle_{w}\vert^{2}}.
\end{equation}
As previously defined, $s=gt/\sigma$, which implies that $s\ll1$
indicates a very short interaction time between the measured system
and the measuring device, resulting in a weak coupling. In this scenario,
after the measurement, the measuring device and the measured system
states remain nearly unchanged from their initially prepared states,
making postselected WM possible. Thus, in this process, the fidelity
between $\vert\phi\rangle$ and $\vert\Psi\rangle$ approaches one,
i.e., $F\to1$, and we say the measurement is weak.

Because the interaction strength between the measured system and the
measuring device is weak during the WM process, the shifts displayed
on the measuring device's dial during a single measurement are insufficient
to achieve the desired outcomes. However, a Zeno-like effect \citep{PhysRevA.41.2295}
can be realized through sufficiently short times or weak coupling,
enabling multiple consecutive measurements to obtain the desired statistical
results. This result is achieved by the signal amplification feature,
which is significant in this measurement domain \citep{2017N,PhysRevA.97.042104,PhysRevA.103.043715}.

Due to the nature of postselected WMs, large weak values are often
accompanied by low postselection probabilities, necessitating multiple
measurements to achieve statistically reliable results. Thus, we conclude
that controlling the coupling strength parameter $s$ to a sufficiently
small value is essential for performing postselected WMs. Our current
study primarily enhances the properties of the TMSV state within this
postselected WM regime.

On the other hand, if we consider strong coupling regimes $(s\gg1)$,
the fidelity function $F$ becomes zero, indicating that our initial
TMSV state $\vert\phi\rangle$ has lost all its properties as it transforms
into a completely different state. In the postselected strong measurement
domain, the weak value loses its significance, and the observable
value of the measured system is interpreted as the postselected (conditional)
expectation value \citep{Aharonov_1991,eq}. In our scheme, this value
of $\hat{\sigma}_{x}$ is given by $\langle\hat{\sigma}_{x}\rangle_{c}=\cos\delta\sin\alpha$
\citep{eq}. In this measurement regime, the initially prepared state
of the measured system is destroyed after a single trial due to the
strong coupling between the measured system and the measuring device,
akin to a strong projective measurement.

In summary, the fidelity analysis between the initial and enhanced
TMSV states reveals that postselected von Neumann measurement can
induce dramatic changes, a feature that can be applied to various
quantum state engineering problems.

\section{Possible implementation methods\label{sec:7} }

In this section, we describe potential implementation methods for
our theoretical scheme in the lab, including ion traps and optical
platforms.

\subsection{ion trap\label{subsec:7-1} }

Based on previous theoretical and experimental works, we introduce
a possible implementation of our postselected von Neumann measurement
in a trapped ion system. As shown in earlier studies \citep{RevModPhys.75.281,RevModPhys.93.025001},
laser-cooled trapped ion systems serve as excellent platforms for
quantum state preparation and manipulation due to their long unwanted
dissipation times relative to the experimental timescales \citep{Hao-Sheng,Nature2020}.
Another advantage of this platform is that the motion of generated
states in trapped ion systems can be completely characterized through
tomographic measurements \citep{PhysRevLett.75.2932,PhysRevLett.77.4281}.
A variety of theoretical schemes have been proposed and experimentally
realized for preparing various motional Gaussian and non-Gaussian
states of trapped ions, including Fock states \citep{PhysRevLett.70.762},
coherent states \citep{PhysRevLett.107.243902,PhysRevA.62.052108,RAFFA2012330,2016Generation},
squeezed states \citep{PhysRevLett.70.556,PhysRevA.52.809}, Schrödinger
cat states \citep{PhysRevLett.76.608,PhysRevA.55.2478}, pair coherent
states \citep{PhysRevA.54.R1014}, pair cat states \citep{PhysRevA.54.4315},
and TMSV states \citep{ZENG2002427,PhysRevA.104.032609}, among others.

In the trapped ion system, we can achieve weak coupling between the
measured system and measuring device using bichromatic light resonant
with the red and blue sidebands. We consider an ion trapped in a harmonic
potential with frequency $\nu$, driven by two laser beams interacting
resonantly with the system tuned to the lower (red) and upper (blue)
vibrational sidebands, respectively. Taking the Lamb-Dicke regime
\citep{111} into account, the total system Hamiltonian in the interaction
picture reads as \citep{PhysRevA.58.761,PhysRevLett.75.2932} 
\begin{align}
H & =\eta\Omega(\hat{\sigma}_{x}\sin\phi_{+}+\hat{\sigma}_{y}\cos\phi_{+})\otimes(\sigma\sin\phi_{-}\hat{P}-\frac{\hbar}{2\sigma}\cos\phi_{-}\hat{X}),\label{eq:62}
\end{align}
where $\eta$ is the Lamb-Dcike parameter, $\text{\ensuremath{\Omega}}$
is the Rabi frequency, and $\phi_{\pm}=\frac{1}{2}\left(\phi_{red}\pm\phi_{blue}\right)$
are the phases associated with the lower and upper sideband laser
phases $\phi_{red}$ and $\phi_{blue}$, respectively. Here, $\hat{X}=\sigma\left(\hat{a}+\hat{a}^{\dagger}\right)$
and $\hat{P}=\frac{i}{2\sigma}\left(\hat{a}^{\dagger}-\hat{a}\right)$
are the position and momentum operators for the external vibrational
part of the ion, and $\sigma=\sqrt{1/2\nu m}$ characterizes the size
of the motional state that depends on the mass $m$ and vibrational
frequency $\nu$ of the ion. Another important point is that we consider
the ion as a two-level system. In the above Hamiltonian, the $\hat{\sigma}_{x}$
and $\hat{\sigma}_{y}$ are the Pauli-$x$ and -$y$ operators, which
we can express in terms of the ion's ground $(\vert\downarrow\rangle)$
and optically excited $(\vert\uparrow\rangle)$ states as $\hat{\sigma}_{x}=\vert\uparrow\rangle\langle\downarrow\vert+\vert\downarrow\rangle\langle\uparrow\vert$
and $\hat{\sigma}_{y}=i\left(\vert\uparrow\rangle\langle\downarrow\vert-\vert\downarrow\rangle\langle\uparrow\vert\right)$,
respectively. The Lamb-Dicke parameter $\eta$ is related to the wave
vector $k$ and defined as $\eta=\frac{k}{\sqrt{2\nu m}}$, and we
assumed $\eta\ll1$.

If we take the external motional states and internal electronic states
of a two-level trapped ion as the measuring device and the measured
system, respectively, Eq. (\ref{eq:62}) describes a typical von Neumann-type
measurement by adjusting relevant parameters. A bichromatic light
field provides their controllable coupling. By setting $\phi_{-}=\frac{\pi}{2}$
and $\phi_{+}=\frac{\pi}{2}$, Eq. (\ref{eq:62}) simplifies to
\begin{equation}
H=g\hat{\sigma}_{x}\otimes\hat{P},\label{eq:63}
\end{equation}
where $g=\eta\sigma\Omega$. This is the Hamiltonian of the von Neumann-type
measurement we introduced in Sec. 2. Using this Hamiltonian, we can
measure the weak value of the observable $\hat{\sigma}_{x}$ in a
trapped ion system. We also note that atomic-system-based WM problems
have been widely studied in the literature \citep{PhysRevA.84.041804,PhysRevLett.111.023604,PhysRevA.92.043825,RN2150,PhysRevA.100.062111,Nature2020Yi,PhysRevA.108.042601,PhysRevA.109.032211}.
Specifically, in experimental works \citep{PhysRevA.100.062111,PhysRevA.108.042601,Nature2020Yi,PhysRevA.109.032211},
the preparation and manipulation of different desired internal states
of a two-level $^{40}Ca^{+}$ trapped ion have been demonstrated.
In the most recent studies \citep{PhysRevA.108.042601,PhysRevA.109.032211},
researchers introduced specific details about the experimental procedures
for implementing the pre- and post-selected measured system states,
which are the same as those used in our theoretical model, in a $^{40}Ca^{+}$
trapped ion system. Since the preparation of external motional states
in trapped ion systems is well-established experimentally, and the
generated states exhibit high stability over long periods, the techniques
employed in these studies could be applied to implement our theoretical
scheme.

\subsection{Optical platform\textcolor{blue}{{} \label{subsec:7-2}}}

\begin{figure}
\begin{centering}
\includegraphics[width=8cm]{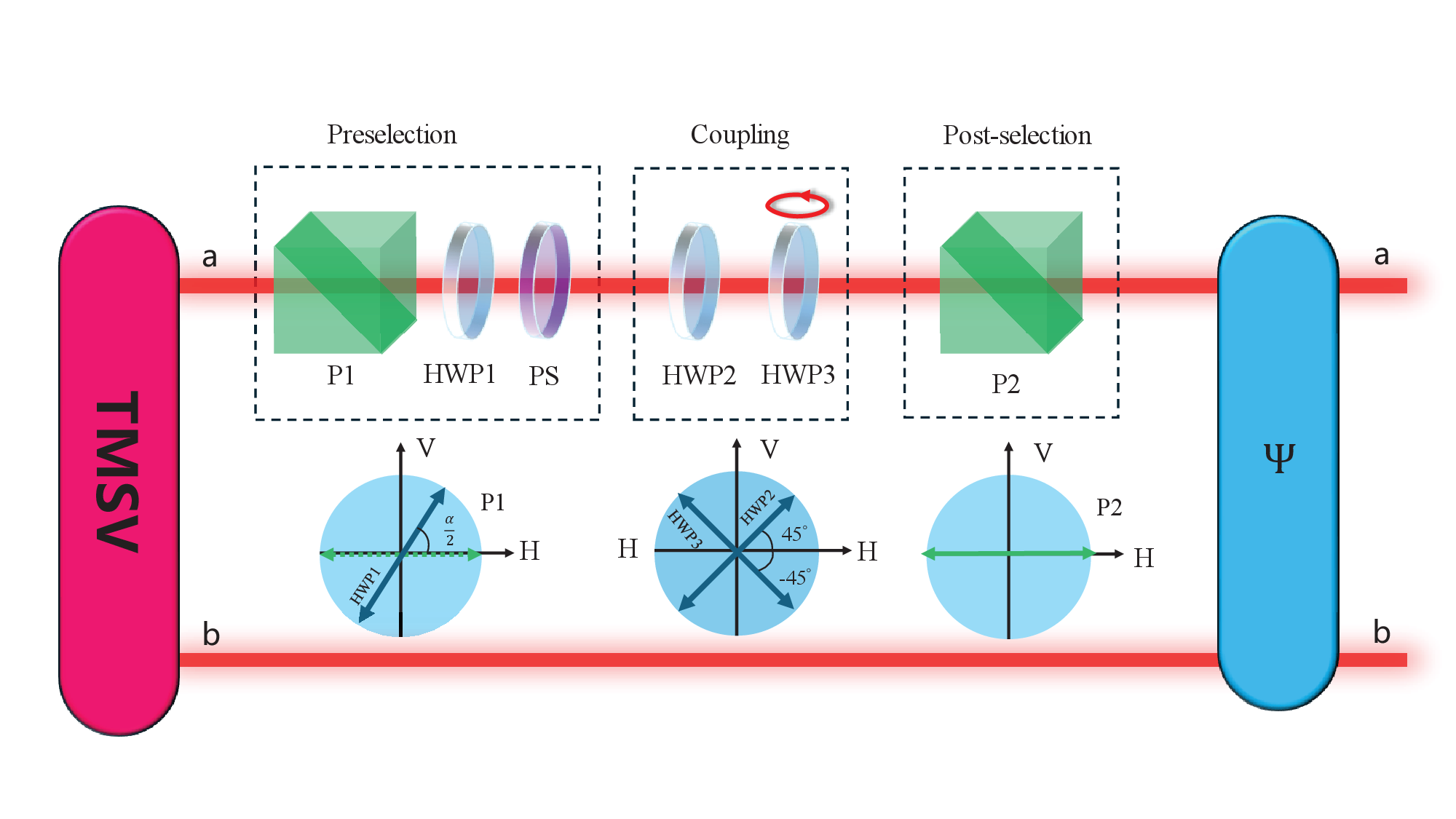}
\par\end{centering}
\caption{\textcolor{blue}{\label{fig:11}}Setup for the realization of the
postselected weak measurement part of our theoretical model using
optical platforms. The TMSV state is generated by optical parametric
down-conversion. We assume that the $a$ mode possesses horizontal
polarization. P1 and P2: Polarizers with optical axes both set at
zero; HWP1: Half waveplate with optical axes set at $\alpha/2$; PS:
Phase shifter used to add a phase shift $\varphi$ to the vertical
polarization. HWP2 and HWP3: Half-waveplate combination with optical
axes perpendicular to each other, with HWP3 rotatable around its optical
axis to introduce a time delay.}
\end{figure}

Our theoretical model can also be realized using optical platforms.
As mentioned in Sec. 2, the TMSV state can be generated by optical
parametric down-conversion \citep{PhysRevResearch.3.033095,Riabinin_2021,REN2019106},
where the two modes correspond to different frequencies of the output
beam, e.g., the signal and idler modes in type-I parametric down-conversion
or different polarizations in type-II parametric down-conversion.
We use the spatial and polarization degrees of freedom of the TMSV
state as the measuring device and measured system, respectively. Fig.
\ref{fig:11} shows the possible experimental setup for implementing
the postselected WM part of our theoretical model via an optical platform.

We assume the TMSV state is generated by optical parametric down-conversion,
with its $a$ mode possessing horizontal polarization. We perform
the postselected WM on the $a$ mode. The initial measured system
state $\vert\psi_{i}\rangle$ can be prepared using a combination
of a linear polarizer (P1), a half-waveplate (HWP1), and a phase shifter
(PS). The P1 polarizer aligns the state to the pure horizontally polarized
state $\vert H\rangle$; HWP1, with optical axes set at $\alpha/2$,
prepares the state $\cos\frac{\alpha}{2}\vert H\rangle+\sin\frac{\alpha}{2}\vert V\rangle$;
and PS adds a phase $\varphi$ to the vertical polarization component,
producing the state $\cos\frac{\alpha}{2}\vert H\rangle+e^{i\varphi}\sin\frac{\alpha}{2}\vert V\rangle$.

Researchers often use birefringent crystals to implement the weak
coupling between the measuring device and the measured system. However,
recent studies \citep{PhysRevLett.111.033604,PhysRevA.103.032212}
have shown that a pair of compound binary zero-order HWPs can replace
the birefringent crystal to introduce time delays between different
polarizations. According to these experimental findings, it is possible
to realize the operator $\hat{\sigma}_{x}=\vert D\rangle\langle D\vert-\vert A\rangle\langle A\vert$
and the weak interaction coupling $g\hat{\sigma}_{x}\hat{P}_{x}$
using two half-waveplates (HWP2 and HWP3), with their optical axes
perpendicular to each other. The time delay is introduced by rotating
HWP3 slightly around its optical axis. Afterward, the second polarizer
(P2) postselects the system into the state $\vert\psi_{f}\rangle=\vert H\rangle$.
Thus, the weak value $\langle\hat{\sigma}_{x}\rangle_{w}$ can be
obtained.

We know that in optical platforms, it is not difficult to investigate
the photon statistics, squeezing, and entanglement properties of TMSV
states using current experimental techniques \citep{PhysRevA.73.042310,PhysRevA.82.021801,PhysRevLett.112.070402,Masa2016,PhysRevA.108.052420}.
Therefore, it may be possible to experimentally verify the theoretical
results presented in this work by utilizing existing optical setups
while incorporating our proposed implementation schemes for postselected
WM (WM).

\section{Discussion \label{sec:8} }

From the fidelity analysis of our measurement output enhanced TMSV
state, we deduced that the postselected von Neumann measurement dramatically
altered the initial TMSV state, even transforming it into a completely
different state. As mentioned in the introduction of this work, the
postselected von Neumann measurement technique is a generalized measurement
framework and can provide alternative approaches to state optimization
problems. Here, by examining the changes in fidelity between the final
and initial TMSV states, we explore potential applications of our
enhanced TMSV state in quantum state engineering (e.g., state preparation
processes). Depending on the coupling strength parameter $s$ and
the squeezing parameter $\lambda$, we introduce its usefulness in
the following four scenarios:

1.Arbitrary values for both the coupling strength parameter $s$ and
the squeezing parameter $\lambda$. Upon carefully examining our state
$\vert\Psi\rangle$ as defined in Eq. (\ref{eq:Psi}), one can observe
that the state $\vert\Psi\rangle$ is a superposition of two distinct
two-mode coherent squeezed vacuum states:
\begin{align}
\vert\Psi\rangle & =\frac{\kappa}{2}\left(t\vert\xi,\frac{s}{2},0\rangle+h\vert\xi,-\frac{s}{2},0\rangle\right)\nonumber \\
 & =\frac{\kappa}{2}\left(t\vert\xi,\frac{s}{2}\rangle+h\vert\xi,-\frac{s}{2}\rangle\right)\vert0\rangle.\label{eq:38-2}
\end{align}
Here, $\vert\xi,\pm\frac{s}{2},0\rangle=D(\pm\frac{s}{2})S(\xi)\vert0,0\rangle=D(\pm\frac{s}{2})\vert\phi\rangle$
with displacement operator $D(\pm\frac{s}{2})$ for the a-mode, and
$t=1+\langle\hat{\sigma}_{x}\rangle_{w}$ and $h=1-\langle\hat{\sigma}_{x}\rangle_{w}$,
respectively. 

2. Strong coupling regime ($s\gg1$) with small squeezing parameter
($\lambda\ll1$). This case corresponds to a postselected strong conditional
measurement. As mentioned in Sec. VI, in this regime, the weak value
loses its significance, and the system observable $\hat{\sigma}_{x}$
is determined by the postselected (conditional) expectation value
$\langle\hat{\sigma}_{x}\rangle_{c}=\cos\delta\sin\alpha$ \citep{eq}.
In this scenario, Eq. (\ref{eq:Psi}) becomes:
\begin{eqnarray}
\vert\Psi^{\prime}\rangle & \approx & \frac{\kappa^{\prime}}{2}\left(t^{\prime}\vert\frac{s}{2}\rangle+h^{\prime}\vert-\frac{s}{2}\rangle\right)\vert0\rangle\nonumber \\
 &  & +\frac{\lambda\kappa^{\prime}}{2}\left(t^{^{\prime}}\vert\frac{s}{2},1\rangle+h^{\prime}\vert-\frac{s}{2},1\rangle\right)\vert1\rangle,
\end{eqnarray}
where $\vert\pm\frac{s}{2},1\rangle$ is displaced Fock state of $a$-mode,
$t^{\prime}=1+\langle\hat{\sigma}_{x}\rangle_{c}$ and $h^{\prime}=1-\langle\hat{\sigma}_{x}\rangle_{c}$,
and 
\begin{equation}
\kappa^{\prime}=\sqrt{2}\left[1+\vert\langle\hat{\sigma_{x}}\rangle_{c}\vert^{2}+(1-\vert\langle\hat{\sigma}_{x}\rangle_{c}\vert^{2})\exp\left(-\frac{s^{2}}{2}\right)\right]^{-\frac{1}{2}}.\label{eq:38-1}
\end{equation}
In this extreme case, the two modes of the TMSV state generate an
entangled state. If the $b$-mode of the TMSV state is in the vacuum
state, the $a$-mode is prepared in a Schrödinger cat-like state $t\vert\frac{s}{2}\rangle+h\vert-\frac{s}{2}\rangle$.
Furthermore, if the $b$-mode contains one photon, then the $a$-mode
remains in a superposition of two displaced Fock states, i.e., $t\vert\frac{s}{2},1\rangle+h\vert-\frac{s}{2},1\rangle$,
though with a very low probability.

3. Coupling strength parameter $s\ll1$ with arbitrary squeezing parameter
$\lambda$. This case belongs to the postselected WM regime, where
the observable values depend on the weak values $\langle\hat{\sigma}_{x}\rangle_{w}$.
In this scenario, Eq. (\ref{eq:38-2}) simplifies to the following
(unnormalized) expression:
\begin{equation}
\vert\Psi^{\prime\prime}\rangle\approx\vert\phi\rangle+\frac{s}{2}Re\left[\langle\sigma_{x}\rangle_{w}\right]\left(a^{\dagger}-a\right)\vert\phi\rangle.\label{eq:39}
\end{equation}
 During the derivation of the above state, we approximate $D\left(\pm\frac{s}{2}\right)\approx\mathbb{I}\pm\frac{s}{2}(\hat{a}^{\dagger}-\hat{a})$,
assuming $s\ll1$, which allows a first-order Taylor expansion. In
this regime, the pointer remains in a superposition of two-mode squeezed
states (TMSS), including single photon-added and single photon-subtracted
TMSV states. Such superposition states may be useful in continuous-variable
quantum teleportation schemes \citep{arora2024}.

For large anomalous weak values $\langle\hat{\sigma}_{x}\rangle_{w}$,
the second term of Eq. (\ref{eq:39}) dominates, contributing significantly
to the signal amplification observed in this work. The enhancement
of the non-classical properties of the TMSV state is achieved for
anomalous weak values $|\langle\hat{\sigma}_{x}\rangle_{w}|\gg1$
in the WM regime ($0<s\ll1$). In this case, the effect of the displacement
operators $\hat{D}\left(\pm\frac{s}{2}\right)$ on the state $\vert\Psi\rangle$
is negligible, and we can approximately obtain TMSSs \citep{PhysRevA.103.062405}.
In this extreme scenario, our results are consistent with those presented
in Ref. \citep{PhysRevA.103.062405} , which discuss odd TMSSs.

Additionally, single-photon-subtracted states have been explored in
quantum state engineering, particularly for generating hybrid entanglement
\citep{RN3} and enhancing entanglement between parties through distillation
processes \citep{RN4}. From the perspective of quantum computation,
single-photon subtraction transforms a Gaussian state into a non-Gaussian
state, facilitating the implementation of universal non-Gaussian gates,
such as the cubic gate \citep{PhysRevA.84.053802,PhysRevA.91.032321}. 

4. Both the coupling strength parameter $s\ll1$ and squeezing parameter
$\lambda\ll1$. In this extreme case, the output state after measurement
is given by the (unnormalized) expression:
\begin{equation}
\vert\Psi^{\prime\prime\prime}\rangle\approx\left(\vert0\rangle+\frac{s}{2}Re\left[\langle\sigma_{x}\rangle_{w}\right]\vert1\rangle\right)\vert0\rangle+\lambda\vert1\rangle\vert1\rangle.\label{eq:40}
\end{equation}
Since $\lambda\ll1$, the $\vert1\rangle\vert1\rangle$ component
can be neglected. We can see that, in this regime, if the $b$-mode
is in the vacuum state, we obtain a superposition of $\vert0\rangle$
and $\vert1\rangle$ for the $a$-mode, or even a single-photon state
of the $a$-mode for sufficiently large weak values. This feature
could be employed in quantum computation processes, as discussed in
\citep{RevModPhys.79.135}. 

However, the most popular method for quantum state engineering in
quantum optics is through the addition or subtraction of single photons
to/from a light field \citep{2007PV}. This optimization and manipulation
technique can be applied to both single and multimode quantum states
of light \citep{2016AV,PhysRevA.110.033717}. It is important to note,
however, that photon addition and subtraction are probabilistic processes,
generally having low success rates. For instance, non-Gaussian optical
states can be generated via heralding schemes using photon detectors,
with photon subtraction typically realized by beam splitters \citep{PhysRevA.91.022317}.
In such cases, transmissivity parameters must be considered when modeling
photon subtraction through a beam splitter, which further reduces
the probability of successful state optimization.

In addition to photon addition and subtraction (or their superpositions),
quantum catalysis is another feasible method for generating nonclassical
quantum states \citep{RN5}. In quantum catalysis, even though no
photons are added or subtracted, it still enhances the quantum properties
of the states because photon catalysis is an inherently quantum mechanical
process. For more details, the reader is referred to \citep{PhysRevLett.96.083601}
and the references therein.

Compared with our current work, previous studies \citep{PhysRevA.95.012310,PhysRevA.103.062405,PhysRevA.103.013705}
have explored the optimization of TMSV states for various purposes.
Specifically, in Ref. \citep{PhysRevA.103.062405}, Cardoso\textsl{
et al.} examined the nonclassical properties of superpositions of
two-mode squeezed states (TMSSs) and introduced methods for preparing
TMSSs by inducing two-mode Jaynes-Cummings and anti-Jaynes-Cummings
interactions in a system involving two modes and a spin-1/2 particle
(within the trapped ion domain). Their results demonstrated that the
nonclassical features of TMSSs, especially for odd TMSSs, can significantly
surpass those of TMSV states under the same parameters. Similarly,
in \citep{PhysRevA.95.012310,PhysRevA.103.013705}, multiphoton catalysis
two-mode squeezed vacuum (MC-TMSV) states were produced using beam
splitters to implement multiphoton catalysis on one or both modes
of a TMSV state. Their numerical results indicated that MC-TMSV states
can significantly improve entanglement, photon statistics, and phase
sensitivity compared to the initial TMSV state. 

As discussed in this paper's numerical analysis, the most significant
enhancement to the properties of the TMSV state via postselected von
Neumann measurement is achieved in the WM regime, particularly with
large anomalous weak values. In this regime, the measurement output
state is characterized with high probability by the second part of
$\vert\Psi^{\prime\prime}\rangle$ defined in Eq. (\ref{eq:39}).
From this perspective, our theoretical proposal can be considered
part of the broader class of existing quantum state manipulation methods.

Compared with previous state optimization methods, this new approach
based on postselected von Neumann measurements can be applied to all
kinds of states, as one can always construct a von Neumann-type interaction
Hamiltonian by selecting two different degrees of freedom of the associated
states. Furthermore, the optimization efficiency and fidelity of the
state can be controlled by adjusting the postselection and anomalous
weak values of the measured system observable within the WM regime.
This work serves as an example of this method.

However, this state optimization technique has its limitations. In
the postselection process, we select only partial information based
on our desired outcome, discarding most information from the initially
prepared state. Another limitation is that large anomalous weak values
are often accompanied by a low probability of successful postselection.
Due to these limitations, achieving the desired results via this method
requires a large number of repeated measurements on the prepared systems,
leading to higher consumption of quantum resources.

The quality of quantum information processing depends on the quality
of the quantum states used as resources. Aside from the potential
applications of our theoretical enhancement proposal of TMSV state
in various quantum information processing scenarios including single
photon generation, advanced imaging techniques, and quantum computation
processes discussed earlier, there are additional practical applications
in the following scenarios: (i) The distillation or enhancement of
entangled states is essential for realizing long-distance quantum
communication \citep{2001P,PhysRevA.78.052312,Xia_2018}. Our enhanced
TMSV state $\vert\Psi\rangle$ can serve as an entangled channel to
improve teleportation fidelity \citep{PhysRevA.95.012310}. (ii) Secure
communication methods require highly sensitive and stable communication
channels between the information sender and receiver. In this context,
the strong correlation characteristics of our enhanced TMSV state
are useful for implementing efficient secure quantum communication,
including quantum cryptography protocols \citep{PhysRevLett.92.057901,Abd-El-Atty2018,2020Pan}.
(iii) It has been shown that postselection operations are advantageous
in quantum metrology \citep{2020Q,PhysRevLett.132.250802} and that
weak value amplification can outperform standard metrology in the
presence of certain types of technical noise \citep{PhysRevX.4.011031}.
Additionally, squeezing and entanglement are critical resources in
precision measurement. Thus, our enhanced TMSV state could benefit
related quantum metrology processes, such as improving phase sensitivity
and increasing the precision of quantum sensors.

In this study, we only considered ideal situations and did not account
for inevitable issues such as decoherence of entanglement \citep{PhysRevResearch.3.033095},
photon losses \citep{PhysRevA.75.053805,PhysRevA.80.013825,PhysRevA.80.063803},
noise \citep{2011,2016} and imperfections in measurement processes
\citep{PhysRevA.86.062325} , all of which can affect the outcomes
of our scheme. In real-world scenarios, systems inevitably interact
with their surrounding environments, leading to decoherence. Consequently,
we may encounter these negative factors during the implementation
of our enhancement proposal for the nonclassical characteristics of
the TMSV state via postselected von Neumann measurement. However,
squeezing is considered an effective method for reducing phase diffusion
\citep{PhysRevA.71.055801,2022P}, and entanglement distillation,
state purification, and concentration methods are valuable for maintaining
entanglement against decoherence \citep{PhysRevLett.76.722,PhysRevA.53.2046,PhysRevA.60.194,2001K,2016Ab}.
Moreover, various techniques exist to avoid or mitigate the detrimental
effects on squeezed states \citep{PhysRevA.78.052312,PhysRevA.81.012325,PhysRevLett.107.083601,2012K,2016W,PhysRevA.105.032609,PhysRevLett.130.073601,2023A,PhysRevA.109.023701,sym16020187}
and two-level systems \citep{PhysRevA.75.012329,PhysRevLett.104.080503,Mangini2022}.
Therefore, the aforementioned results not only help to maintain the
nonclassical properties of our output state $\vert\Psi\rangle$ but
also may ensure the experimental feasibility of postselected WMs under
real-world conditions. Further details regarding the real-world experimental
implications of this work are beyond the scope of the current research
and remain an open problem for future studies.

\section{Conclusion and outlooks\label{sec:9}}

In summary, we have investigated the effects of postselected von Neumann
measurement characterized by postselection and weak values on the
nonclassical features of the TMSV state. The nonclassicality of our
pointer (measuring device) state was analyzed through calculations
of quadrature squeezing, second-order correlation functions, and various
entanglement criteria. We found that all these quantities, which describe
the quantumness of the objective state, can be enhanced in the WM
regime with anomalous weak values. This implies a constructive role
for postselected WM in improving the nonclassical nature of the TMSV
state. It is well known that the state $\vert\phi\rangle$ is a Gaussian
entangled state. However, our results indicate that the enhanced TMSV
state can be considered a non-Gaussian entangled state resulting from
the postselected von Neumann measurement applied to the single-mode
($a$ mode) of the initial TMSV state. Since the nonzero quantum correlation
$\langle\hat{a}\,\hat{b}\rangle$ is a crucial property of the TMSV
state, the enhancements proposed in our work can be attributed to
the positive effects of postselected measurement on $\langle\hat{a}\,\hat{b}\rangle$. 

In terms of quadrature squeezing, the initial TMSV state $\vert\phi\rangle$
exhibits squeezing in only one quadrature. However, we found that
after applying postselected measurement, squeezing effects can occur
in both quadratures of the TMSV in the WM regime for large weak values,
within certain regions of the squeezing parameter $\lambda$. By investigating
the quantum statistics of our enhanced TMSV state using the Mandel-$Q$
parameter and second-order correlation functions, we demonstrated
that postselected WM positively influences the quantumness of our
considered state. Specifically, after postselection, the $a$ mode
of the TMSV state transitioned from super-Poissonian to sub-Poissonian,
while the $b$ mode exhibited a superbunching effect in regions of
small squeezing parameters. Additionally, we examined the entanglement
properties of our measurement output state by employing measures such
as the concurrence, the HZC criterion, and EPR correlations. We observed
significant improvements in all correlations in WM regions with large
weak values. Notably, the enhanced regions of HZC and EPR correlations
covered most parameter regions, indicating that these measures provide
a better representation for assessing the entanglement of our state.
To clarify the results of this paper further, we also analyzed the
fidelity between the initial TMSV state $\vert\phi\rangle$ and the
enhanced TMSV state $\vert\Psi\rangle$. 

Finally, we proposed potential implementations of our theoretical
scheme in optical and ion trap platforms. We introduced the postselected
WM procedures and discussed their feasibility in laboratory settings.
We also briefly compared the advantages and disadvantages of our scheme
to existing state optimization methods and highlighted its potential
applications in related quantum information processes. In our current
work, we explored the possibility of performing postselected von Neumann
measurements in two-mode radiation fields. Future research could focus
on the effects of postselected von Neumann measurements on other Gaussian
and non-Gaussian multipartite continuous-variable radiation fields
\citep{2020quantum,PRXQuantum.2.030204,PhysRevA.109.040101}, including
two-photon coherent states \citep{Agarwal:88,PhysRevA.50.2865,PhysRevA.13.2226},
three-mode states \citep{1990,PhysRevA.64.052303}, and other multimode
radiation fields \citep{PhysRevA.42.4102}.
\begin{acknowledgments}
This work was supported by the National Natural Science Foundation
of China (No. 12365005) .
\end{acknowledgments}

\appendix

\section{\label{sec:A1}Related expression}

In this paper, we derive explicit expressions for the relevant average
values associated with our proposed state $\vert\Psi\rangle$. However,
due to their complexity, many of these expressions are too cumbersome
to include in the main text. Therefore, we have provided them in this
appendix for reference. \begin{widetext} 

\begin{align}
\langle\hat{a}\rangle & =s\vert\kappa\vert^{2}\left\{ \frac{1}{2}\Re\left[\langle\hat{\sigma_{x}}\rangle_{w}\right]-i\Im\left[\langle\hat{\sigma_{x}}\rangle_{w}\right]\left[\cosh^{2}\lambda-\frac{1}{2}\right]\beta\right\} ,\\
\langle\hat{b}\rangle & =\frac{s\vert\kappa\vert^{2}}{2}\left\{ \Re\left[\langle\hat{\sigma_{x}}\rangle_{w}\right]+i\Im\left[\langle\hat{\sigma_{x}}\rangle_{w}\right]\left[\sinh\left(2\lambda\right)+1\right]\beta\right\} ,\\
\langle\hat{a}^{2}\rangle & =\frac{s^{2}\vert\kappa\vert^{2}}{8}\left(1+\vert\langle\hat{\sigma_{x}}\rangle_{w}\vert^{2}\right)\nonumber \\
 & \quad+\frac{s^{2}\vert\kappa\vert^{2}}{2}\left(1-\vert\langle\hat{\sigma_{x}}\rangle_{w}\vert^{2}\right)\left[\cosh^{4}\lambda-\cosh^{2}\lambda+\frac{1}{4}\right]\beta,\\
\langle\hat{b}^{2}\rangle & =\frac{s^{2}\vert\kappa\vert^{2}}{8}\left(1+\vert\langle\hat{\sigma_{x}}\rangle_{w}\vert^{2}\right)\nonumber \\
 & \quad+\frac{s^{2}\vert\kappa\vert^{2}}{4}\left(1-\vert\langle\hat{\sigma_{x}}\rangle_{w}\vert^{2}\right)\left[\sinh^{2}\left(2\lambda\right)+\sinh\left(2\lambda\right)+\frac{1}{2}\right]\beta\\
\langle\hat{a}^{\dagger}\hat{a}\rangle & =\frac{\vert\kappa\vert^{2}}{2}\left(1+\vert\langle\hat{\sigma_{x}}\rangle_{w}\vert^{2}\right)(\sinh^{2}\lambda+\frac{s^{2}}{4})\nonumber \\
 & \quad+\frac{\vert\kappa\vert^{2}}{2}\left(1-\vert\langle\hat{\sigma_{x}}\rangle_{w}\vert^{2}\right)\left(\sinh^{2}\lambda\left(1-s^{2}\cosh^{2}\lambda\right)-\frac{s^{2}}{4}\right)\beta,\\
\langle\hat{b}^{\dagger}\hat{b}\rangle & =\frac{\vert\kappa\vert^{2}}{2}\left(1+\vert\langle\hat{\sigma_{x}}\rangle_{w}\vert^{2}\right)\left(\sinh^{2}\lambda+\frac{s^{2}}{4}\right)\nonumber \\
 & \quad+\frac{\vert\kappa\vert^{2}}{2}\left(1-\vert\langle\hat{\sigma_{x}}\rangle_{w}\vert^{2}\right)\left[\sinh^{2}\lambda\left(1-s^{2}\cosh^{2}\lambda\right)+\frac{s^{2}}{4}\right]\beta
\end{align}
\begin{align}
\langle\hat{a}^{\dagger}\hat{b}\rangle & =\frac{s^{2}\vert\kappa\vert^{2}}{8}\left(1+\vert\langle\hat{\sigma_{x}}\rangle_{w}\vert^{2}\right)+\frac{s^{2}\vert\kappa\vert^{2}}{2}\left(1-\vert\langle\hat{\sigma_{x}}\rangle_{w}\vert^{2}\right)\nonumber \\
 & \quad\times\left[\sinh^{3}\lambda\cosh\lambda+\frac{1}{2}(\sinh^{2}\lambda+\frac{1}{2}\sinh\left(2\lambda\right))+\frac{1}{4}\right]\beta,\\
\langle\hat{a}\hat{b}\rangle & =\frac{\vert\kappa\vert^{2}}{4}\{+\left(1-\vert\langle\hat{\sigma_{x}}\rangle_{w}\vert^{2}\right)\left[\sinh\left(2\lambda\right)\left(1-s^{2}\cosh^{2}\lambda\right)-s^{2}\left(\cosh^{2}\lambda-\frac{1}{2}\sinh\left(2\lambda\right)+\frac{1}{2}\right)\right]\beta\nonumber \\
 & \quad+\left(1+\vert\langle\hat{\sigma_{x}}\rangle_{w}\vert^{2}\right)(\sinh\left(2\lambda\right)+\frac{s^{2}}{2})\},\\
\left\langle \hat{a}^{\dagger}\hat{a}\hat{b}^{\dagger}\hat{b}\right\rangle  & =\frac{\vert\kappa\vert^{2}}{2}\{\left(1+\vert\langle\hat{\sigma_{x}}\rangle_{w}\vert^{2}\right)k_{0}+\beta\left(1-\vert\langle\hat{\sigma_{x}}\rangle_{w}\vert^{2}\right)[k_{1}-\frac{s}{2}\left(k_{2}+k_{3}+k_{4}+k_{5}\right)\nonumber \\
 & \quad-\frac{s^{4}}{16}+\frac{s^{2}}{4}\left[2\sinh^{2}\lambda\left(1-s^{2}\cosh^{2}\lambda\right)+\sinh\left(2\lambda\right)\right]]\},\\
\langle\hat{a}^{\dagger2}\hat{a}^{2}\rangle & =\frac{\vert\kappa\vert^{2}}{2}\{\left(1+\vert\langle\hat{\sigma_{x}}\rangle_{w}\vert^{2}\right)T_{0}+\beta\left(1-\vert\langle\hat{\sigma_{x}}\rangle_{w}\vert^{2}\right)[T_{1}-s(T_{2}+T_{3})\nonumber \\
 & \quad+\frac{s^{2}}{4}(s^{2}\sinh^{4}\lambda+s^{2}\cosh^{4}\lambda+4\sinh^{2}\lambda\left(1-s^{2}\cosh^{2}\lambda\right))-\frac{3s^{4}}{16}]\},\\
\langle\hat{b}^{\dagger2}\hat{b}^{2}\rangle & =\frac{\vert\kappa\vert^{2}}{2}\{\left(1+\vert\langle\hat{\sigma_{x}}\rangle_{w}\vert^{2}\right)T_{0}+\beta\left(1-\vert\langle\hat{\sigma_{x}}\rangle_{w}\vert^{2}\right)[T_{1}-s(T_{4}+T_{3})\nonumber \\
 & \quad+\frac{s^{2}}{2}(\frac{s^{2}}{4}\sinh^{2}(2\lambda)+2\sinh^{2}\lambda\left(1-s^{2}\cosh^{2}\lambda\right))+\frac{s^{4}}{16}]\}.
\end{align}
 Here, \begin{subequations} 
\begin{equation}
k_{0}=\sinh^{2}\lambda\cosh(2\lambda)+\frac{s^{4}}{16}+\frac{s^{2}}{2}\sinh\lambda(\sinh\lambda+\cosh\lambda),
\end{equation}
\begin{equation}
k_{1}=\sinh^{2}\lambda\cosh(2\lambda)+\frac{s^{2}\sinh^{2}(2\lambda)}{4}\left(\frac{s^{2}}{4}\sinh^{2}(2\lambda)-4\sinh^{2}\lambda-1\right),
\end{equation}
\begin{equation}
k_{2}=\frac{s\sinh(2\lambda)}{8}\left(4\cosh(2\lambda)-s^{2}\sinh^{2}(2\lambda)\right),
\end{equation}
\begin{equation}
k_{3}=\frac{s\sinh(2\lambda)\sinh^{2}\lambda}{2}\left(s^{2}\cosh^{2}\lambda-2\right),
\end{equation}
\begin{equation}
k_{4}=\sinh^{2}\lambda\left(s^{2}\sinh^{2}\lambda\cosh^{2}\lambda-\cosh(2\lambda)\right),
\end{equation}
\begin{equation}
k_{5}=\frac{s\sinh^{2}(2\lambda)}{4}\left(2-s^{2}\cosh^{2}\lambda\right),
\end{equation}
\begin{equation}
T_{0}=2\sinh^{4}\lambda+s^{2}\sinh^{2}\lambda+\frac{s^{4}}{16},
\end{equation}
\begin{equation}
T_{1}=s^{2}\sinh^{4}\lambda\cosh^{2}\lambda(s^{2}\cosh^{2}\lambda-4)+2\sinh^{4}\lambda,
\end{equation}
\begin{equation}
T_{2}=s\sinh^{4}\lambda\left(s^{2}\cosh^{2}\lambda-2\right),
\end{equation}
\begin{equation}
T_{3}=\frac{s}{4}\sinh^{2}(2\lambda)\left(s^{2}\sinh^{2}\lambda+2\right),
\end{equation}
\begin{equation}
T_{4}=s\cosh\lambda\sinh^{3}\lambda\left(2-s^{2}\cosh^{2}\lambda\right).
\end{equation}

\end{subequations}\end{widetext}

\bibliographystyle{apsrev4-1}
\bibliography{ref-tow-one}

\begin{thebibliography}{174}
\bibitem [{eq()}]{eq}%
  \BibitemOpen
  \href@noop {} {}\bibinfo {note} {If the eigvenvalue fucntion of system observable $\hat{A}$ is $\hat{A}\vert a_{i}\rangle=a_{i}\vert a_{i}\rangle$,then conditioanl expectation value of $\hat{A}$ for preselected state $\vert\psi_{i}\rangle$ and postselected state $\vert\psi_{f}\rangle$ can be written as\[\langle\hat{A}\rangle_{c}=\frac{\sum_{i}a_{i}\vert\langle\psi_{f}\vert a_{i}\rangle\langle a_{i}\vert\psi_{i}\rangle\vert^{2}}{\sum_{j}\vert\langle\psi_{f}\vert a_{j}\rangle\langle a_{j}\vert\psi_{i}\rangle\vert^{2}}\]}\BibitemShut {NoStop}%
\end{thebibliography}

\end{document}